\documentclass[useAMS,usenatbib]{mn2e}
\input{psfig}
\newcommand{\etal}{{et al.~}}

\newcommand{\kmsmpc}{\>{\rm km}\,{\rm s}^{-1}\,{\rm Mpc}^{-1}}

\newcommand{\Msun}{\>{\rm M_{\odot}}}

\newcommand{\Msunh}{\>h^{-1}\rm M_\odot}

\newcommand{\beq}{\begin{equation}}
\newcommand{\eeq}{\end{equation}}

\newcommand{\calC}{{\cal C}}

\newcommand{\Mh}{M_{\rm h}}

\def\gtsima{$\; \buildrel > \over \sim \;$}
\def\ltsima{$\; \buildrel < \over \sim \;$}
\def\prosima{$\; \buildrel \propto \over \sim \;$}
\def\gsim{\lower.7ex\hbox{\gtsima}}
\def\lsim{\lower.7ex\hbox{\ltsima}}
\def\simgt{\lower.7ex\hbox{\gtsima}}
\def\simlt{\lower.7ex\hbox{\ltsima}}
\def\simpr{\lower.7ex\hbox{\prosima}}
\def\la{\lsim}
\def\ga{\gsim}
\def\lta{\la}
\def\gta{\ga}

\newcommand{\apj}{ApJ}

\newcommand{\aj}{AJ}
\newcommand{\mnras}{MNRAS}

\newdimen\hssize
\newdimen\hdsize
\begin{document}


\title[Galaxy Activity in Dark-Matter Haloes]
      {The Rise and Fall of Galaxy Activity in Dark Matter Haloes}

\author[A. Pasquali et al.]
       {\parbox[t]{\textwidth}{
        Anna Pasquali$^{1}$\thanks{E-mail:pasquali@mpia.de}, 
        Frank C. van den Bosch$^{1}$, 
        H.J. Mo$^{2}$,
        Xiaohu Yang$^{3,4}$, \\
        Rachel Somerville$^{1,5}$} 
        \vspace*{10pt} \\
  $^{1}$Max-Planck Institut f\"ur Astronomie, K\"onigstuhl 17, 
        69117 Heidelberg, Germany\\
  $^{2}$Department of Astronomy, University of Massachusetts, 
        Amherst, MA 01003-9305, USA\\
  $^{3}$Shanghai Astronomical Observatory, The Partner Group of MPA, 
        Nandan Road 80, Shanghai 200030, China\\
  $^{4}$Joint Institute for Galaxy and Cosmology (JOINGC) of Shanghai 
        Astronomical Observatory and University of Science\\
        and Technology of China\\
  $^{5}$Space Telescope Science Institute, 3700 San Martin Drive,
        Baltimore, MD 21218, USA}


\date{}
\pagerange{\pageref{firstpage}--\pageref{lastpage}}
\pubyear{2008}

\maketitle

\label{firstpage}


\begin{abstract}
  We  use the catalogue  of  galaxy groups constructed  from the Sloan
  Digital Sky  Survey (SDSS DR4) by Yang  et al.  (2007)  to study the
  dependence  of galaxy activity  on   stellar mass, $M_{\ast}$,  halo
  mass, $M_{\rm  h}$ and group hierarchy  (central {\it  vs} satellite
  galaxies).  The wealth of  data provided by the   SDSS allows us  to
  split  the  sample on the basis  of  galaxy activity in star-forming
  galaxies, galaxies  with  optical AGN activity,   composite galaxies
  (both star formation  and optical AGN  activity) and  radio sources.
  We find a smooth transition in halo  mass as the activity of central
  galaxies  changes from star   formation to  optical AGN  activity to
  radio emission.   Star-forming  centrals  preferentially  reside  in
  haloes  with $M_{\rm h}   < 10^{12} \Msunh$,  central  galaxies with
  optical-AGN activity typically inhabit haloes  with $M_{\rm h}  \sim
  10^{13} \Msunh$, and centrals emitting in the radio mainly reside in
  haloes more  massive than $10^{14} \Msunh$.   Although this seems to
  suggest that the  environment (halo  mass)  determines the type   of
  activity of its central galaxy, we find a similar trend with stellar
  mass:  central star formers typically  have stellar masses less than
  $10^{10}  h^{-2}\Msun$,  while optical-AGN  hosts  and central radio
  sources have characteristic stellar masses of $\sim 10^{10.8} h^{-2}
  \Msun$ and $\sim 10^{11.6}  h^{-2} \Msun$, respectively.  Since more
  massive haloes typically  host more massive  centrals, it is unclear
  whether the  activity of a  central galaxy  is causally connected to
  its stellar mass    or to its  halo  mass.   In   general, satellite
  galaxies have  their  activity  suppressed with  respect  to central
  galaxies  of  the same stellar mass.   This  holds not only for star
  formation activity, but also for AGN activity in the optical and the
  radio. At fixed stellar mass, we find that the activity of satellite
  galaxies depends only weakly on  halo mass. In fact,  using a set of
  reduced conditional probability     functions, we   find  that   for
  satellite galaxies the dependence of galaxy activity on halo mass is
  more than four times weaker than the dependence on stellar mass.  We
  also investigate  whether  the   strength  of  galaxy  activity   is
  influenced  by the  environment, using the  specific luminosities in
  H$\alpha$, [OIII] and at    1.4 GHz.  We    find that the   specific
  luminosity of  star-forming and  optical-AGN centrals systematically
  decreases with  increasing halo mass,  while the specific luminosity
  of centrals  with  high  radio activity  increases with  halo  mass.
  Independently  of the activity   class,  the specific  luminosity of
  satellite galaxies does not significantly  vary with halo mass.  All
  these results are consistent with a picture in which low mass haloes
  accrete cold gas, while massive haloes have coronae  of hot gas that
  promote radio activity of their central galaxies.
\end{abstract}




\clearpage
\section{Introduction}
\label{sec:intro}

Ever since  the seminal work  of Hubble (1926),  it is known  that the
local  galaxy  population  consists  of  two main  branches:  the  red
sequence  populated  by  bulge-dominated  galaxies  with  old  stellar
populations (early-type galaxies), and  the blue cloud comprising disk
galaxies  with  some   level  of  star-formation  activity  (late-type
galaxies). With the help of large galaxy redshift surveys, such as the
Sloan  Digital Sky  Survey  (SDSS; York  \etal  2000; Stoughton  \etal
2002), this bimodality has  been dramatically confirmed (e.g., Blanton
et  al.  2003;  Strateva et  al.  2001).   This bimodality  is tightly
linked  to environment,  with early-types  preferentially  residing in
dense environments  such as  galaxy clusters, while  late-types mainly
populate  low-density structures such  as galaxy  groups and  the more
generic  field (e.g.,  Dressler 1980;  Postman \&  Geller  1984; Lewis
\etal 2002;  Goto \etal 2003;  Kauffmann \etal 2004; Hogg  \etal 2004;
Weinmann \etal 2006).

In  the current  paradigm of  galaxy  formation, it  is believed  that
virtually all  galaxies initially form as late-type  disk galaxies due
to the  cooling of  gas with non-zero  angular momentum  in virialized
dark  matter  haloes  (e.g.,  White  \&  Frenk  1991).   During  their
subsequent   evolution,  disk   galaxies  may   be   transformed  into
early-types via a variety  of transformation mechanisms.  Probably the
most important of  these is major mergers, in  which two disk galaxies
of roughly equal mass merge  to produce a spheroidal galaxy (Toomre \&
Toomre 1972; Negroponte \& White  1983).  During such a merger the gas
looses its angular momentum and accumulates at the center where it can
trigger an intense  starburst and fuel the central  black hole (e.g.,
Hernquist 1989; Mihos  \& Hernquist 1996).  This not  only couples the
growth of the central black hole to that of its host galaxy, which may
explain the observed $M_{\rm BH}$ - $\sigma$ relation (Gebhardt et al.
2000; Ferrarese  \& Merritt 2000),  but it may also  quench subsequent
star  formation by  exhausting  the  gas supply  and  by expelling  or
heating  the  cold gas  via  AGN  activity  (e.g., Menci  \etal  2005;
Springel,  Di  Matteo \&  Hernquist  2005;  Hopkins  \etal 2006).   In
addition  to this  so-called quasar  mode  of AGN  feedback, one  also
considers  the  so-called radio  mode,  in  which  the radio  activity
associated with a  low accretion efficiency of the  central black hole
provides  a  heating  term  that  can offset  the  cooling,  therewith
preventing any  new gas  from cooling and  forming stars. It  has been
suggested  that  this radio-mode  feedback  plays  a  crucial role  in
explaining why the most massive galaxies are red and dead (e.g., Bower
\etal 2006; Croton \etal 2006; Cattaneo \etal 2006; Kang, Jing \& Silk
2006;  Nusser, Silk  \& Babul  2006; Sijacki  \& Springel  2006).  The
inclusion  of AGN  feedback  (radio mode  and/or  quasar mode)  allows
models of galaxy formation to reproduce the bright end of the observed
luminosity  function. In  order  to  reproduce the  faint  end of  the
luminosity function, the models typically invoke supernova feedback to
expel  gas from low  mass haloes  (e.g., Larson  1974; White  \& Rees
1978; Dekel \& Silk 1986).

Clearly, star formation and AGN activity are thought to play a pivotal
role in the formation and evolution of galaxies. An important, largely
outstanding  question,  is  how  this  activity is  influenced  by  the
galaxy's environment. Such  a link will put strong  constraints on the
mechanisms  responsible for  triggering and  quenching such  activity. 
For example, if AGN were to be fueled by the same cold gas that feeds
star formation in disk  galaxies, their environmental dependence would
most  probably follow  the  one of  star-forming  galaxies, i.e.   the
fraction of AGN would decrease in denser environments.  Alternatively,
if the AGN activity were maintained by galaxy-galaxy interactions, it
would be preferentially associated  with intermediate mass groups with
low  velocity dispersions  or with  young  clusters that  are not  yet
virialized (e.g., Gunn 1979; Shlosman et al.  1990). 

The  existence, in  the local  Universe,  of a  star-formation rate  -
density relation is well established  by now.  Hashimoto et al. (1998)
measured the environment in terms  of local density, which is based on
the distance of  a galaxy to its third nearest  neighbor. They found a
continuous  correlation between  star-formation rate  (SFR)  and local
density, whereby  galaxies with a higher SFR  reside preferentially in
lower density  environments. In addition, the authors  showed that low
levels of  star formation  depend more strongly  on the  local density
within  a cluster  than  in the  field,  while high  SFRs (which  they
identified as starbursts) are less  sensitive to local density in both
clusters and the field. This  suggest that two processes determine the
SFR - density relation: gas removal via stripping, responsible for the
low  SFRs  of  galaxies   in  dense  environments,  and  galaxy-galaxy
interactions,    causing    the    prevalence   of    starbursts    in
intermediate-density  environments. Similar  results were  obtained by
Carter et  al. (2001)  and G\'omez  et al. (2003)  who found  that the
fraction  of star-forming  galaxies decreases  with  increasing galaxy
density (derived  from the  distance of a  galaxy to its  10th nearest
neighbor). In particular, Lewis et al. (2002) and G\'omez et al. found
a break  in this trend  at a critical  density of 1  h$^{-2}$ Mpc$^2$,
below which the fraction of star-forming galaxies is high and constant
and above which it drops rapidly.  The authors also found that the SFR
increases with the cluster-centric  radius, to reach the SFR typically
measured in  the field at  radii larger than  3 - 4 times  the cluster
virial radius.  They concluded that this trend can not be explained by
the  morphology  -  density  relation  alone,  against  the  intuitive
interpretation of  the SFR - density  relation as a  direct product of
the environmental  dependence of galaxy morphology.  Kauffmann  et al. 
(2004) found  a  decrease in  the  specific  star-formation rate  (i.e
SFR/M$_*$)  of   a  factor  of   10  between  low-   and  high-density
environments.   More  recently,  Weinmann  et  al.   (2006),  using  a
catalogue of  galaxy groups  from the Sloan  Digital Sky  Survey (SDSS
DR2), have  been able to quantify  the galaxy environment  in terms of
halo  mass, i.e. the  mass of  the dark-matter  halo hosting  a galaxy
group.  The authors have shown that the fraction of late-type galaxies
(actively forming stars) smoothly increases with decreasing halo mass,
down to  10$^{12}$ M$_{\odot}$.   For any given  halo mass,  this same
fraction increases with the halo-centric radius.

As  for AGN,  a  number of  studies  in the  literature  have found  a
constant fraction of  optical AGN at low redshift  across a wide range
of  environments, from the  field to  galaxy clusters  (Monaco et  al. 
1994; Coziol et al. 1998;  Shimada et al.  2000; Schmitt 2001; G\'omez
et al.  2003;  Miller et al.  2003). Kauffmann  et al. (2004) obtained
rather  opposite  results,  indicating  that the  fraction  (at  fixed
stellar  mass)   of  AGN  with  strong  [OIII]   emission  doubles  in
low-density environments  with respect to  the denser ones.   In their
study,  environment is primarily  based on  the number  of neighboring
galaxies, but  a comparison with cosmological  simulations allowed the
authors   to  assign   to   the  lowest/highest   local  densities   a
characteristic  halo mass  of  10$^{12}$ -  10$^{13}$ M$_{\odot}$  and
10$^{14}$ - 10$^{15}$ M$_{\odot}$, respectively.

A different approach to studying the environment dependence of optical
AGN activity has been used by Wake et al. (2004), Croom et al.  (2005)
and  Li et  al.  (2006),  who measured  the  cross-correlation between
optical AGN and  a reference sample of inactive  galaxies. The authors
agree  that  optical  AGN  and  the  control  sample  share  the  same
clustering amplitude  on scales  larger than a  few Mpc. In  the range
between 100  kpc and  1 Mpc, however,  optical AGN are  clustered more
weakly than the control galaxies, while for scales smaller than 70 kpc
there is  a weak indication for  optical AGN to  cluster more strongly
than inactive  galaxies.  Using N-body simulations, Li  et al.  reveal
that the AGN  anti-bias measured between 100 kpc and  1 Mpc is possibly
due to AGN residing preferentially  at the center of their dark-matter
haloes.  In addition, the fraction of optical AGN located in 10$^{12}$
- 10$^{13}$  M$_{\odot}$ haloes  is found  to be  larger than  that in
haloes as massive as 10$^{14}$ - 10$^{15}$ M$_{\odot}$.

Based on  their [OIII]  luminosity, optical AGN  are often  divided in
LINERs and Seyferts, where  the latter have higher [OIII] luminosities
and accretion rates (e.g., Kewley  et al.  2006).  Kelm et al.  (2004)
and  Constantin \&  Vogeley (2006)  have  shown that  LINERs are  more
strongly clustered  than Seyferts.  This trend is  consistent with the
morphology-density relation, given that LINERs are generally hosted by
more massive galaxies (cf.  Kewley et al.  2006).  However, Constantin
et al.   (2007) found that  LINERs associated with  moderately massive
host galaxies ($<$ 3 $\times$ 10$^{10}$ M$_{\odot}$) appear to be more
common  in voids (the  most underdense  environments in  the Universe)
than  in  walls,  although  their  accretion  rates  and  old  stellar
populations are comparable to their wall counterparts. Void AGN hosted
by more massive  galaxies seem, instead, to accrete  more strongly and
to have younger stellar populations than those in walls.  These trends
would thus suggest that the AGN activity in voids is tightly connected
with  the  availability of  fuel,  the  same  fuel that  supports  the
star-formation activity of their host galaxies.

On the basis  of their radio activity, AGN  are commonly distinguished
between  radio-quiet and  radio-loud. These  two groups  are  known to
exhibit different clustering properties, which are also different from
those of optical AGN.  For  example, Yee \& Green (1984) and Ellingson
et al. (1991) found that radio-loud AGN are located in galaxy clusters
as  rich as  Abell class  0/1, possibly  reflecting the  picture where
radio-loud AGN  reside in  elliptical galaxies, while  radio-quiet AGN
are mostly associated with disk  galaxies.  On the contrary, McLure \&
Dunlop (2001) found that  the clustering amplitudes of radio-quiet and
radio-loud AGN  are rather similar, and  both groups of  radio AGN are
detected in environments as rich  as Abell class 0. Later on, Prestage
\& Peacock (1988), Hill \& Lilly (1991), Miller et al. (2002) and Best
(2004) claimed that  radio-loud AGN appear to favor  galaxy groups and
weak clusters  and to avoid  the densest environments unless  they are
hosted by  the central  cluster galaxy.  A  more in-depth  analysis by
Best  (2004)  has shown  that  the  fraction  of radio-loud  AGN  with
absorption  lines  (in  their  optical spectra)  increases  in  denser
environments while the fraction  of radio-loud AGN with emission lines
decreases.  Significantly  larger samples of  radio AGN such  as those
drawn from the  SDSS survey have confirmed the  early results obtained
by Best (2004) and have  indicated that radio-loud AGN are more likely
hosted by the brightest group or cluster galaxies (Best et al.  2005a,
2007; Kauffmann, Heckman \& Best 2008).

In   the  studies   mentioned  above,   the  environment   is  usually
parameterized  in terms  of a  projected number  density  of galaxies,
above  a  given  magnitude  limit.   Typically  this  number  density,
indicated by  $\Sigma_n$, is measured using the  projected distance to
the  $n$th nearest neighbor,  with $n$  typically in  the range  5-10. 
However,  as  discussed  in  Weinmann  et al.   (2006),  the  physical
interpretation  of  $\Sigma_n$   itself  depends  on  environment:  in
clusters,  where the  number  of  galaxies is  much  larger than  $n$,
$\Sigma_n$  measures  a  {\it   local}  number  density,  which  is  a
sub-property of  the cluster (i.e., $\Sigma_n$  is strongly correlated
with cluster-centric radius).   In low-density environments, which are
populated by haloes which typically  contain only one or two galaxies,
$\Sigma_n$ measures a much more  global density, covering a scale that
is  much larger  than  the halo  in  which the  galaxy resides.   This
ambiguous,  physical  meaning  of  $\Sigma_n$ severely  complicates  a
proper interpretation of  the various correlations between environment
and galaxy properties. Furthermore, this environment indicator can not
distinguish   between  central  and   satellite  galaxies.    This  is
important, since models of galaxy formation and evolution predict that
these two  types of galaxies interact  with their host  haloes in very
different  ways: while  central  galaxies can  accrete  halo gas  that
cools, and  may drive subsequent  star formation and/or  AGN activity,
satellite  galaxies  are  subjected   to  various  processes  such  as
strangulation,  ram-pressure stripping  and  galaxy harassment  (Bower
2006; Croton  et al.  2006;  Kang, Jing \&  Silk 2006; Hopkins et  al. 
2006;  Larson, Tinsley  \& Caldwell  1980; Balogh,  Navarro  \& Morris
2000; Gunn \& Gott 1972; Farouki \& Shapiro 1981; Moore et al.  1996).

In  this paper,  we re-visit  the environmental  dependence  of galaxy
activity (star  formation, optical  AGN activity and  radio emission),
using a large galaxy-group  catalogue constructed from the SDSS.  This
catalogue allows  us to parameterize  the environment in terms  of the
halo  mass  in which  each  galaxy resides  and  to  split the  galaxy
population  in  centrals and  satellites,  thus  allowing  for a  more
straightforward   comparison  with   galaxy   formation  models.    In
particular, the  observational results presented  here place important
constraints on  the various mechanisms responsible  for triggering and
quenching  different  sort  of  activity  in  central  and  satellites
galaxies.   Throughout  this  paper   we  adopt  a  flat  $\Lambda$CDM
cosmology with $\Omega_{\rm m} = 0.238$ and $\Omega_{\Lambda} = 0.762$
(Spergel et al.  2007) and we  express units that depend on the Hubble
constant in terms of $h = H_0/(100 \kmsmpc)$.

\section{Data}
\label{sec:data}

The analysis presented  in this paper is based on  the SDSS DR4 galaxy
group  catalogue of  Yang  \etal (2007;  hereafter  Y07).  This  group
catalogue is constructed applying  the halo-based group finder of Yang
\etal (2005a) to the  New York University Value-Added Galaxy Catalogue
(NYU-VAGC;  see  Blanton \etal  2005),  which  is  based on  SDSS  DR4
(Adelman-McCarthy \etal  2006).  From this catalogue  Y07 selected all
galaxies  in  the Main  Galaxy  Sample  with  an extinction  corrected
apparent magnitude  brighter than $r=18$, with redshifts  in the range
$0.01 \leq  z \leq 0.20$ and  with a redshift  completeness $\calC_z >
0.7$.   This sample  of  galaxies  is used  to  construct three  group
samples: Sample~I, which only uses the $362356$ galaxies with measured
redshifts from the SDSS, Sample~II which also includes $7091$ galaxies
with  SDSS  photometry  but  with  redshifts  taken  from  alternative
surveys, and Sample~III  which includes an additional $38672$ galaxies
that  lack a  redshift due  to fiber-collisions,  but which  have been
assigned  the redshift of  their nearest  neighbor (cf.   Zehavi \etal
2002).  Since the assignment of  optical activity to the galaxies (see
below)  requires  SDSS  spectra,  the  present analysis  is  based  on
Sample~I.

The magnitudes and  colours of all galaxies are  based on the standard
SDSS Petrosian  technique (Petrosian  1976; Strauss \etal  2002), have
been corrected for galactic  extinction (Schlegel, Finkbeiner \& Davis
1998), and have been $K$-corrected and evolution corrected to $z=0.1$,
using  the method  described  in  Blanton \etal  (2003).   We use  the
notation $^{0.1}M_r$ to indicate  the resulting absolute magnitudes in
the $r$-band. Stellar masses  for all galaxies, denoted by $M_{\ast}$,
are computed  using the relations between  stellar mass-to-light ratio
and colour of Bell \etal (2003; see Y07 for details).

As described in Y07, the majority  of the groups in our catalogue have
two estimates of  their dark matter halo mass $\Mh$:  one based on the
ranking of its total characteristic luminosity, and the other based on
the ranking  of its  total characteristic stellar  mass.  As  shown in
Y07, both halo masses agree very well with each other, with an average
scatter that  decreases from  $\sim 0.1$  dex at the  low mass  end to
$\sim 0.05$ dex at the  massive end.  In addition, detailed tests with
mock  galaxy redshift  catalogues have  demonstrated that  these group
masses are more  reliable than those based on  the velocity dispersion
of the group members (Yang \etal 2005b; Weinmann \etal 2006; Y07).  In
this  paper  we adopt  the  group masses  based  on  the stellar  mass
ranking\footnote{We have  verified, though,  that none of  our results
  changes significantly  if we  adopt the luminosity-rank  based masses
  instead.}.   These masses  are  available for  a  total of  $228062$
groups in  our sample,  which host a  total of $290729$  galaxies that
constitute  the  sample of  galaxies  that  we  use for  our  analysis
described  below.   

Note  that this  is not  a volume-limited  sample.   Consequently, the
sample suffers from Malmquist  bias, causing an artificial increase of
the average  luminosity (and also  stellar mass) of the  galaxies with
increasing redshift.  To correct for  this bias, we weight each galaxy
by $1/V_{\rm max}$, where $V_{\rm  max}$ is the comoving volume of the
Universe out  to a comoving distance  at which the  galaxy would still
have made the  selection criteria of our sample.   In what follows all
distributions  are weighted  by $1/V_{\rm  max}$,  unless specifically
stated otherwise.

Using  the group  catalogue, we  split our  sample  into ``centrals'',
which are defined as the most  massive group members in terms of their
stellar mass,  and ``satellites'', which are those  group members that
are  not centrals.   In what  follows,  we refer  to this  distinction
between centrals and satellites  as the ``group hierarchy''.  Finally,
for each group member we determine the projected radius $R_{\rm proj}$
from the luminosity weighted group center using the angular separation
and the  angular diameter distance at  the redshift of  the group.  We
normalize these  projected radii by  the characteristic radius  of the
group, $R_{180}$, which  is defined as the radius  inside of which the
dark matter halo associated with  the group has an average overdensity
of $180$.  This  radius is computed from the mass  and redshift of the
group using equation~(5) in Y07.

\begin{figure}
\centerline{\psfig{figure=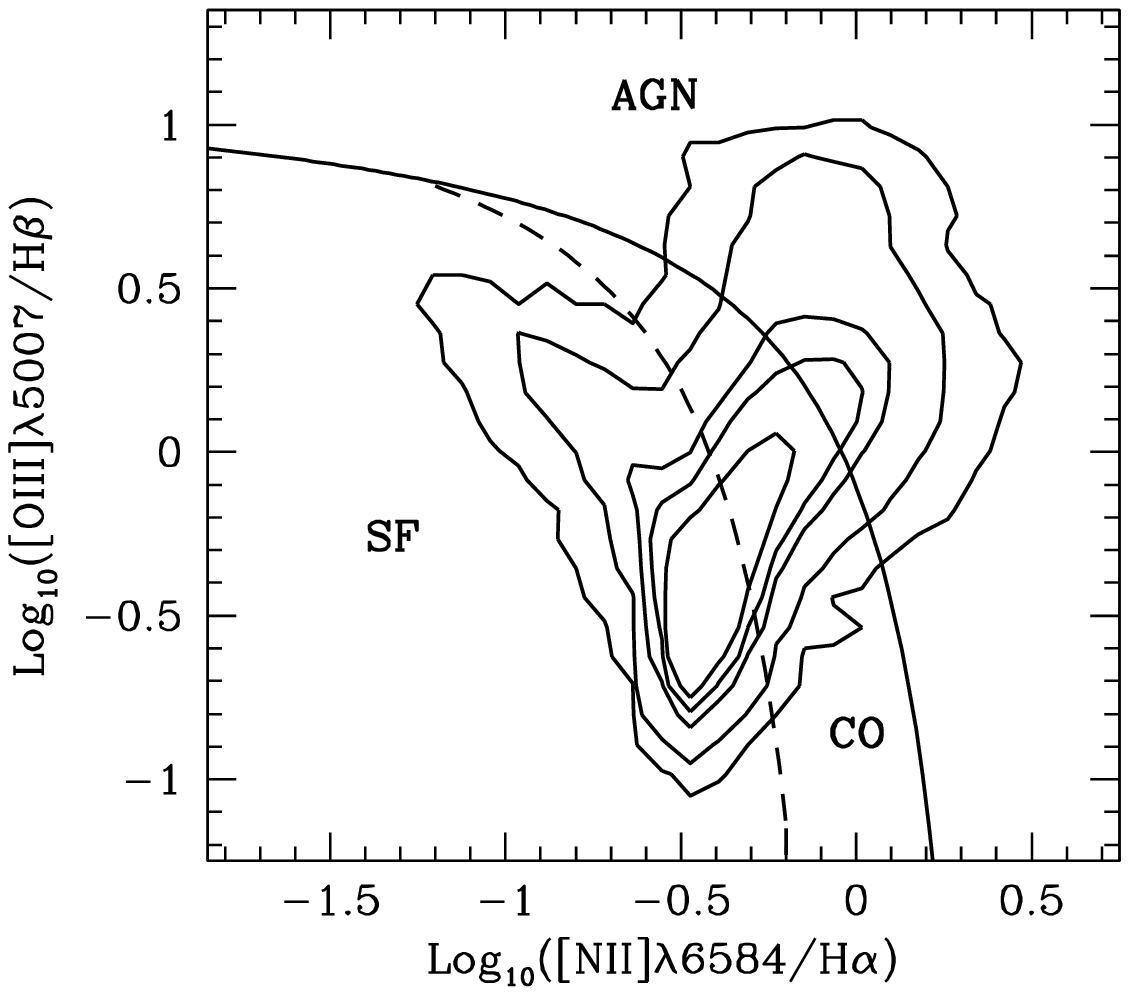,width=\hssize}}
\caption{The BPT diagram (Baldwin, Philips  \& Terlevich 1981) for the
  galaxies  in  our   sample  whose  [OIII]$\lambda$5007 and H$\alpha$
  emission lines have a S/N ratio $\geq$ 3.   The black contours trace
  2D bins (of  $0.1$ dex $\times  0.1$ dex) containing an equal number
  of galaxies (contour levels: 3, 50, 500, 1000 and 2000).  The dashed
  line represents the demarcation  line of pure star formation defined
  by   Kauffmann \etal  (2003b) and the    solid  line is  the extreme
  starburst demarcation line of Kewley \etal (2001).}
\label{fig:bpt}
\end{figure}

\subsection{Optical Activity}
\label{sec:optact}

In order to  assign the galaxies in our sample to  one of four classes
of  optical activity, we  match the  galaxies in  our sample  with the
catalogue  of  emission-line  fluxes  constructed by  Kauffmann  \etal
(2003b). These  authors measured various nebular fluxes  from the SDSS
spectra  corrected for  Galactic extinction  after subtraction  of the
stellar  absorption-line spectrum.   The latter  is fitted  with model
galaxy spectra  computed using the stellar population  code of Bruzual
\&  Charlot  (2003),  and   the  best-fitting  synthetic  spectrum  is
subtracted from the observed  spectrum.  Emission-line fluxes are then
corrected for the intrinsic reddening  from the galaxy itself using an
extinction curve  of the form $\lambda^{-0.7}$ (Charlot  \& Fall 2000)
and assuming the canonical  flux ratio H$\alpha$/H$\beta$ = 2.86 (Case
I, Osterbrock 1989).  

\begin{figure*}
\centerline{\psfig{figure=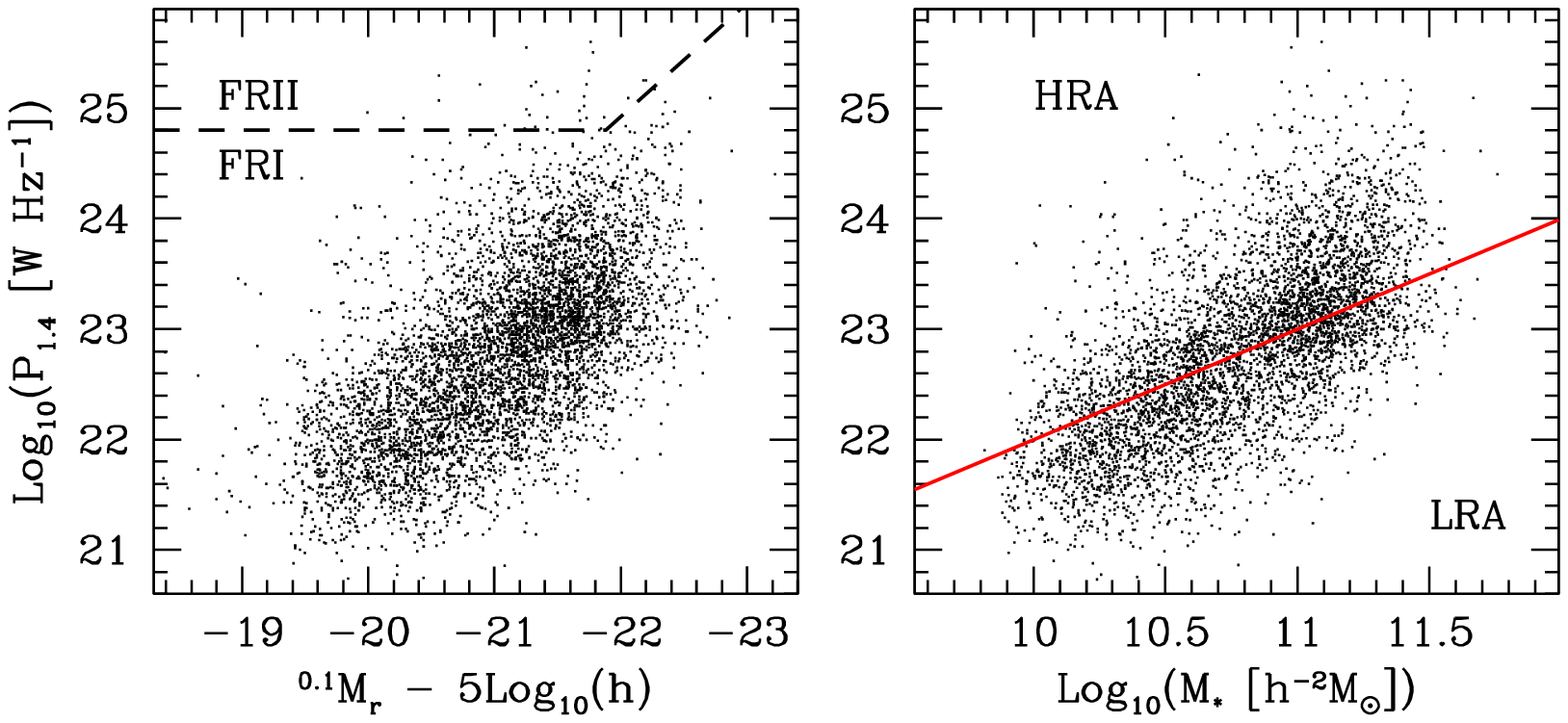,width=12cm}}
\caption{{\it Left-hand panel:} the power at 1.4 GHz, $P_{1.4}$,
  (in W Hz$^{-1}$)  of the radio sources in our  sample that reside in
  groups  with an assigned  halo mass  as a  function of  the absolute
  magnitude,  $^{0.1}M_r -  5\log(h)$,  of their  host galaxies.   The
  dashed  lines  indicate the  location  of  FRII-type objects.   {\it
    Right-hand panel:}  $P_{1.4}$ as a function of  the galaxy stellar
  mass, $M_{\ast}$.  The solid line corresponds to $P_{1.4}/M_{\ast} =
  10^{12}$ W~Hz$^{-1}/(h^{-2}$M$_{\odot}^{-1}$)  and separates sources
  with high  radio activity (HRA, $P_{1.4}/M_* >  10^{12}$) from those
  with low radio activity (LRA, $P_{1.4}/M_* \leq 10^{12}$).}
\label{fig:radiopower}
\end{figure*}

We   consider   a   galaxy   to   be   `optically   active'   if   its
[OIII]$\lambda$5007 and  H$\alpha$ emission lines are  detected with a
S/N ratio  $\geq$ 3. In  order to discriminate  between star-formation
and     AGN    activity,     we    use     the     line-flux    ratios
[OIII]$\lambda$5007/H$\beta$ and [NII]$\lambda$6584/H$\alpha$ to place
the galaxies  in our  sample on the  BPT diagram (Baldwin,  Philips \&
Terlevich 1981).  This diagram  is traditionally used to separate type
II  AGN from  star-forming  galaxies and  composite  galaxies. In  the
standard unified  model (e.g.  Antonucci 1993), AGN  are classified as
type I if their central  black hole and its broad emission-line region
are viewed directly. When, instead, the central black hole is obscured
and only its narrow emission-line  region is visible, AGN are labeled
as  type II.  Since  the photometry  and  spectra of  type  I AGN  are
dominated by non-thermal  emission, they can not be  used to study the
properties of their host galaxies, such as stellar mass. Consequently,
AGN of type I are excluded from our sample.

Fig.~\ref{fig:bpt}  shows  the BPT  diagram  of  the  galaxies in  our
sample. The dashed line represents the
demarcation  line of pure  star formation  defined by  Kauffmann \etal
(2003b, their equation 1) and  the solid line is the extreme starburst
demarcation line derived by Kewley \etal (2001, their equation 5).  We
classify as star-forming (SF) all galaxies that in the BPT diagram lie
below the  pure star formation  line and as  type II AGN  all galaxies
above the  extreme-starburst line.  Those galaxies in  between the two
lines  are labeled  as composites  (CO).  The  remaining  galaxies in
sample I  that either do  not satisfy our  S/N requirements or  do not
exhibit emission lines are classified as `not optically active' (NOA).
The number  of star-forming galaxies  (SF), optical AGN  and composite
(CO) galaxies  is listed  in Table~1, where  we also  indicate whether
they are centrals or satellites.

The  label AGN  comprises both  Seyfert  II galaxies  and LINERs  (the
latter   have   a    lower   [OIII]$\lambda$5007/H$\beta$   at   fixed
[NII]$\lambda$6584/H$\alpha$).  Kewley  \etal (2006) have demonstrated
that Seyferts of type II  and LINERs form a continuous sequence, where
LINERs   have   a   lower    accretion   rate   than   Seyferts.    In
\S\ref{sec:strength}  we will  use the  specific [OIII]  luminosity to
split our sample in type II Seyferts and LINERS, but for the main part
of this paper we do not discriminate between these two types of AGN.

\begin{table}
\centering
\caption{Overview of Sample}
\begin{tabular}{lrr}
\hline
    & Centrals & Satellites \\
\hline
ALL &   228062 & 62667 \\
SF  &    51738 & 18017 \\
CO  &    35174 &  7358 \\
AGN &    24273 &  4506 \\
HRA &     2595 &   390 \\
LRA &     2594 &   366 \\
\hline
\end{tabular}
\medskip
\begin{minipage}{\hssize}
  {\it Notes:}  The numbers of  centrals and satellites in  our sample
  and in  the various classes  of optical and radio  activity (SF=star
  forming,  CO=composite,  AGN=optical  AGN  activity,  LRA=low  radio
  activity, HRA=high radio activity).
\end{minipage}
\end{table}

\subsection{Radio Activity}
\label{sec:radact}

In  addition to  the  optical  activity, we  also  consider the  radio
activity of the  galaxies in our sample.  To that  extent we match our
galaxy  sample   with  the  NVSS  survey   (National  Radio  Astronomy
Observatory Very Large  Array Sky Survey; Condon et  al. 1998) and the
FIRST survey  (Faint Images  of the Radio  Sky at  Twenty centimeters;
Becker et  al.  1995), both conducted  at a frequency of  1.4 GHz.  We
follow  the identification  procedure of  Best \etal  (2005b),  and we
refer the reader to their paper for details.  Contrary to Best et al.,
who only selected galaxies with a flux density $\geq$ 5 mJy, we select
all galaxies detected at 1.4 GHz, independent of their flux densities.
Using the  observed flux density and  the SDSS redshift  we derive the
radio power  at 1.4  GHz, $P_{1.4}$,  which we plot  as a  function of
galaxy absolute magnitude in  the $^{0.1}r$-band in the left-hand side
panel  of Fig.  2.   Here, the  dashed lines  mark the  area typically
populated by FRII-type objects according to Owen \& White (1991).  The
FRI/FRII  classification was  first  introduced by  Fanaroff \&  Riley
(1974): in FRI sources the radio emission peaks near the center of the
galaxy  and the  emission  from the  jets  fades with  galacto-centric
distance.  FRII sources, on the other hand, have edge-brightened radio
lobes.  As  is evident from  Fig.~\ref{fig:radiopower}, the population
of  radio galaxies  in  the  local Universe  is  clearly dominated  by
FRI-type objects,  which are  typically hosted by  elliptical galaxies
with  very little  on-going  star formation  and  weak emission  lines
(Ledlow \&  Owen 1995; Govoni \etal  2000; Best \etal  2005b).  In the
right-hand panel of Fig. 2 we plot $P_{1.4}$ of our sample galaxies as
a function of  their stellar mass $M_*$; the  solid line indicates the
locus  where the specific  radio power,  $P_{1.4}/M_* =  10^{12} h^{2}
{\rm W}  {\rm Hz}^{-1} M_{\odot}^{-1}$, which coincides  with the peak
of the corresponding  distribution (cf. Fig. 15).  We  split our radio
galaxies in sources with low  radio activity (LRA) and those with high
radio activity (HRA), depending on whether $P_{1.4}/M_*$ is smaller or
larger  than $10^{12}  h^{2}  {\rm W}  {\rm Hz}^{-1}  M_{\odot}^{-1}$,
respectively.  The numbers  of LRA and HRA galaxies  identified in our
sample, and split according  to centrals and satellites, are indicated
in Table~1.

The  optical  and  radio   activity  defined  here  are  not  mutually
exclusive.  In  fact, it  is well known  that radio galaxies  can also
display optical activity. This overlap is summarized in Table~2, where
we  list  the  number  of  LRA  and  HRA  centrals  and
satellites with optical  activity (SF, CO or AGN)  and with no optical
activity  (NOA).   Note that  the  majority  (75  percent) of  the LRA 
galaxies also reveal optical activity, roughly equally split among SF,
CO and  AGN. However, in the case  of the HRA galaxies  50 percent does
not reveal  any optical  activity. Forty percent  of HRA  galaxies with
optical activity  are identified  as AGN, while  only 26  percent show
evidence for ongoing star formation.
\begin{table}
\centering
\caption{Optical Activity of Radio Galaxies.}
\begin{tabular}{lrrrr}
\hline
               &  SF  &  CO & AGN & NOA \\
\hline
LRA centrals    &  685 & 606 & 653 &  651\\
LRA satellites  &  149 &  91 &  79 &   71\\
HRA centrals    &  322 & 426 & 538 & 1308\\
HRA satellites  &   66 &  61 &  57 &  182\\
\hline
\end{tabular}
\medskip
\begin{minipage}{\hssize}
  {\it  Notes:} The numbers  of LRA and HRA 
  centrals  and satellites  that are  also identified  as star-forming
  galaxies (SF), composite galaxies (CO), optical AGN, or that are not
  optically active  (NOA). 
\end{minipage}
\end{table}

\section{The Ecology of Galaxy Activity}
\label{sec:eco}

Using the sample defined above, we now investigate how the optical and
radio activity of galaxies correlate  with their stellar mass and with
the mass of the dark matter halo in which they reside. In addition, we
also examine whether the group hierarchy (central {\it vs.} satellite)
has any impact  on the presence of activity in  the optical and/or the
radio.
\begin{figure*}
\centerline{\psfig{figure=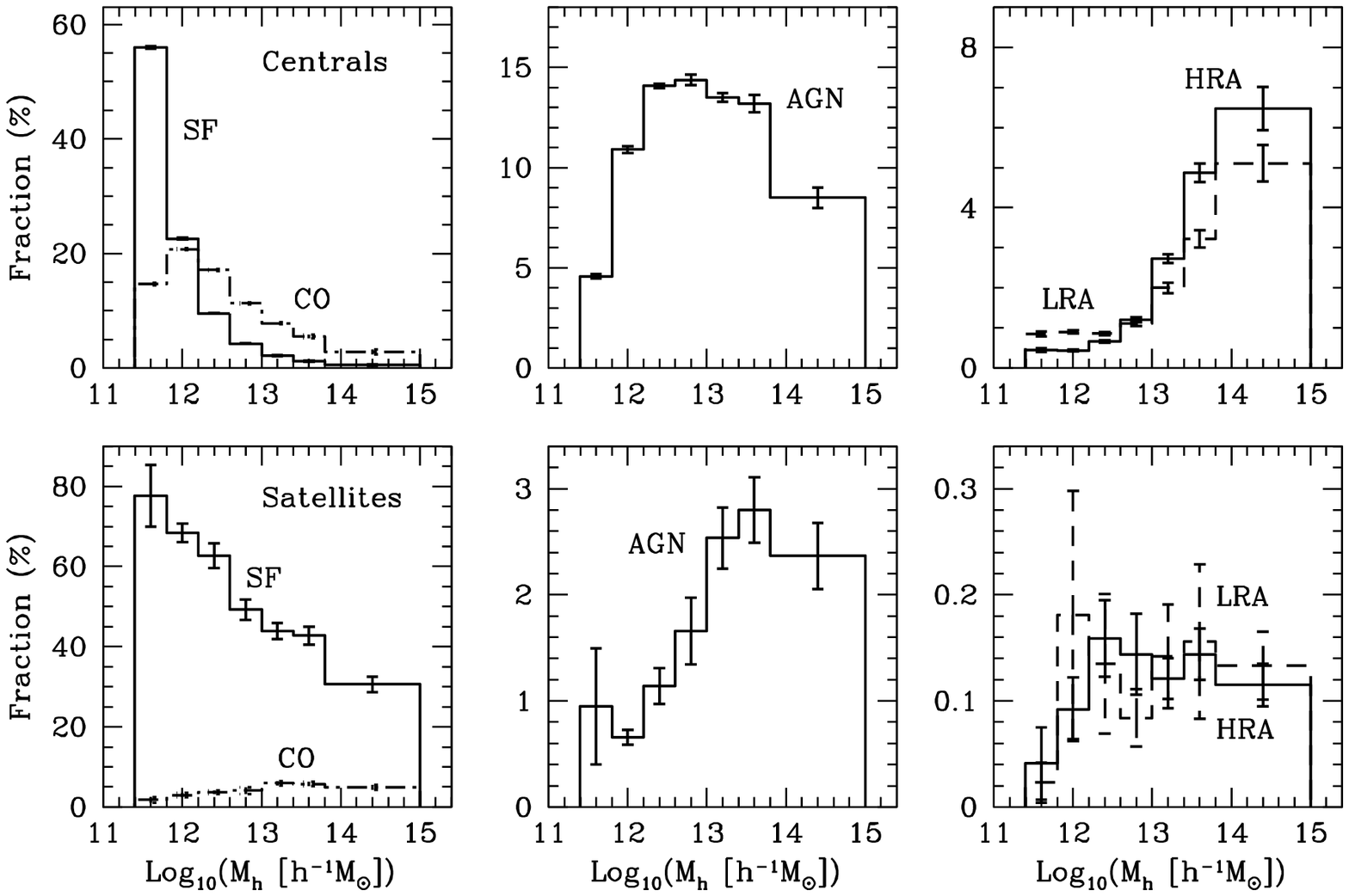,width=16cm}}
\caption{The activity fractions $f(A|M_{\rm h},{\rm C})$ for centrals 
  (upper panels)  and $f(A|M_{\rm h},{\rm S})$  for satellite galaxies
  (lower panels), as functions of  their halo mass, $M_{\rm h}$. These
  fractions  reflect  the  halo  mass  distributions  of  the  various
  activity  classes  (SF,  CO,  AGN,  LRA and  HRA,  as  labeled).  The
  errorbars  have  been  computed   using  the  jackknife  method,  as
  described in the text, and correspond to 1$\sigma$.}
\label{fig:halomass}
\end{figure*}

We define  $f(A|P)$ as  the fraction of  galaxies with  properties $P$
(halo mass, stellar mass, group  hierarchy) that are in activity class
$A$ (SF, CO, AGN, LRA, HRA). These fractions are computed using
\begin{equation}
f(A|P) = \sum\limits_{i=1}^{N_{A|P}} w_i /
         \sum\limits_{i=1}^{N_P} w_i\,.
\end{equation}
Here $w_i = 1/V_{{\rm max},i}$ is the weight of galaxy $i$, defined as
the  reciprocal  of the  comoving  volume of  the  Universe  out to  a
comoving  distance at  which  the  galaxy would  still  have made  the
selection criteria of our sample,  $N_{A|P}$ is the number of galaxies
with properties $P$  that are in activity class $A$,  and $N_P$ is the
total number  of galaxies with properties $P$.   Errors are determined
using  the jackknife technique.   We divide  the group  catalogue into
$N=20$ subsamples  of roughly equal size, and  recalculate $f(A|P)$ 20
times, each time leaving out  one of the 20 subsamples.  The jackknife
estimate of the standard deviation then follows from
\begin{equation}
\label{jack}
\sigma_f = \sqrt{{N-1 \over N} \sum_{i=1}^N \left(f_i - \bar{f}\right)^2}
\end{equation}
with  $f_i$  the fraction  obtained  from  jackknife  sample $i$,  and
$\bar{f}$ the average.

\subsection{Halo Mass Dependence}
\label{sec:halomass}

The upper and lower panels of Fig.~\ref{fig:halomass} show $f(A|M_{\rm
  h},{\rm C})$  and $f(A|M_{\rm h},{\rm S})$,  respectively.  Here `C'
and `S'  refer to  centrals and satellites,  respectively, $A$  is the
activity class  (SF, CO,  AGN, LRA, HRA), and $M_{\rm  h}$ is  the halo
mass.   Thus, these  plots show  the  halo mass  distributions of  the
various activity classes, split  according to centrals and satellites. 
Note how  the peak  of the halo  mass distribution of  centrals shifts
towards  a  higher halo  mass  as  their  activity changes  from  star
formation  to optical  AGN activity  to radio  emission.  Star-forming
centrals prefer  the least  massive environments with  a halo  mass of
$\sim 10^{11.6} h^{-1} \Msun$ (this is the minimum halo mass probed by
the galaxy group catalogue of  Y07); their fraction drops rapidly with
increasing halo  mass, and becomes  negligible in haloes with  $M \gta
10^{14}  h^{-1}  \Msun$.  Central  galaxies  classified as  composites
prefer somewhat more massive  haloes; the halo mass distribution peaks
around $\sim 10^{12} h^{-1} \Msun$ and remains non-negligible up to the
scale  of massive  clusters.  The  fraction of  centrals  harboring an
optically  classified AGN rises  sharply from  $M_{\rm h}  = 10^{11.6}
h^{-1} \Msun$ to $10^{12.2}  h^{-1}\Msun$, stays roughly constant at a
level of about 14\% out  to $M_{\rm h} \sim 10^{13.8} h^{-1}\Msun$ and
decreases by  nearly a factor of  2 at the massive  end.  Finally, the
percentage of  central radio  galaxies increases with  increasing halo
mass, and  reaches a maximum in  the most massive halo  mass bin.
The results  for HRA and LRA  centrals are remarkably  similar, with a
weak trend for the HRA centrals to prefer more massive haloes than LRA
centrals. Thus, when moving to more massive haloes, the characteristic
activity of  central galaxies changes  from star formation  to optical
AGN  activity to  radio  activity.  Note,  though,  that the  absolute
fraction  of  `active'   galaxies  decreases  quite  drastically  with
increasing halo  mass; while more than  60 percent of  the centrals in
low mass  haloes ($M_{\rm h}  \sim 10^{11.6} h^{-1}\Msun$)  are active
(mostly star  forming), at the massive  end only $\sim  20$ percent of
the  centrals are  active,  split roughly  equally  among optical  AGN
activity and radio activity (both HRA and LRA).

With  regard  to  satellite  galaxies,  the  fraction  of  those  with
star-formation activity diminishes  by a factor of 3  going from group
masses of $\sim 10^{11.6} h^{-1}\Msun$ to $10^{15} h^{-1}\Msun$.  Note
that in massive  haloes, the fraction of satellite  galaxies with star
formation  activity is  an  order  of magnitude  higher  than that  of
central galaxies. The fractions of CO and AGN satellites increase with
increasing halo mass. Especially  for the composites, this behavior is
quite different from that of  the centrals.  Finally, the fractions of
HRA and  LRA satellites  is rather constant  across the
full range of halo masses probed. Note, though, that only a very small
fraction of satellite galaxies  ($\lta 0.2$ percent) are radio active,
which  is  roughly  an  order  of magnitude  lower  than  for  central
galaxies.

\begin{figure*}
\centerline{\psfig{figure=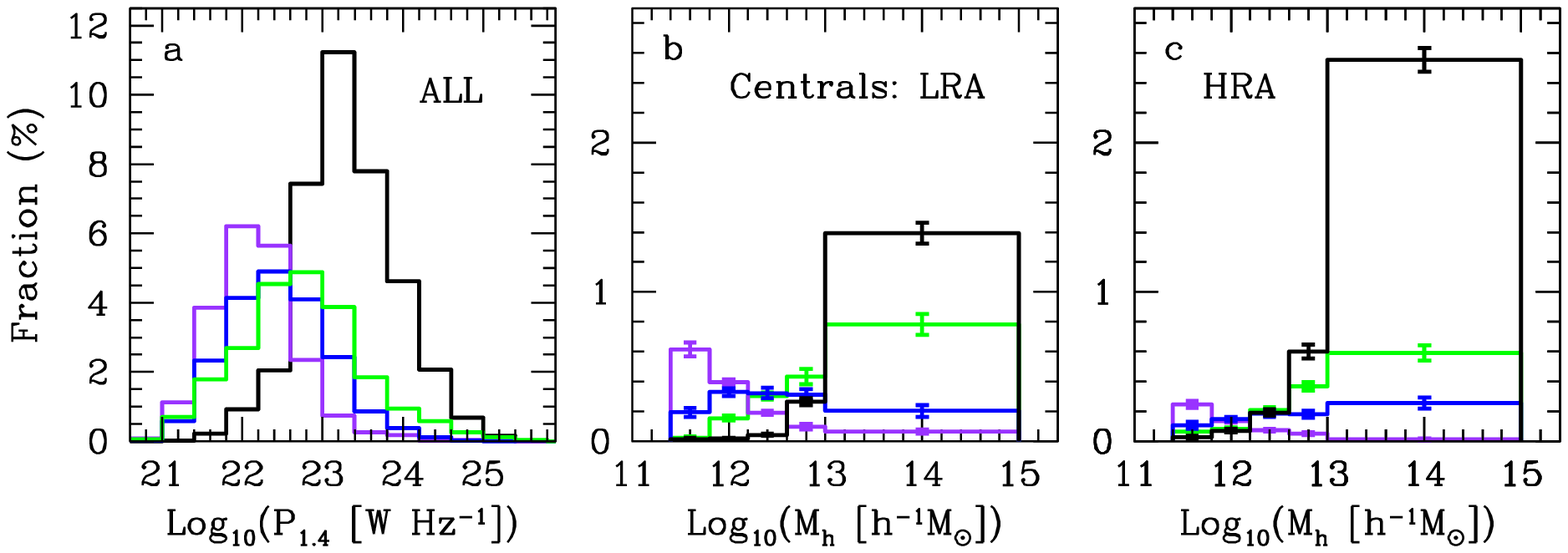,width=15cm}}
\caption{{\it Panel a:} the distribution in radio power of all sources,
  distinguished among  star-forming galaxies (purple  line), composite
  galaxies (blue line), AGN hosts  (green line) and radio sources with
  no optical  activity (black line).  The fractions on the  y-axis are
  computed with  respect to the total  number of radio  sources in our
  sample. {\it  Panel b:} the distribution  of radio-emitting centrals
  with low radio activity (LRA)  as function of halo mass, split among
  star-forming  galaxies, composite  galaxies, AGN  hosts  and sources
  with no optical activity (the colour coding is as in panel {\it a}).
  The fractions on the y-axis are calculated with respect to the total
  number  of central galaxies  in our  sample.  {\it  Panel c:}  as in
  panel  {\it b},  but for  radio  centrals with  high radio  activity
  (HRA).}
\label{fig:radiodetail}
\end{figure*}

As discussed  in Section~\ref{sec:radact}, optical  and radio activity
are   not   mutually  exclusive   (cf.    Table~2).    Panel  (a)   of
Fig.~\ref{fig:radiodetail} shows the  distribution of $P_{1.4}$ of all
radio sources for each optical activity class.  Here, the fractions on
the  y-axis are computed  with respect  to the  total number  of radio
galaxies  in  our sample.   SF  galaxies  dominate  at $P_{1.4}  \leq$
10$^{22.6}$ WHz$^{-1}$,  while the AGN distribution  extends to larger
values, in agreement with the  findings of Best (2004).  The NOA radio
sources  are, instead, preferentially  associated with  $P_{1.4} \geq$
10$^{23}$ WHz$^{-1}$.  Panels (b) and (c) show the fractions $f(A_{\rm
  opt}+A_{\rm   radio}|M_{\rm  h},{\rm   C})$,  where   $A_{\rm  opt}$
corresponds  to an optical  activity class  (SF, CO,  AGN or  NOA) and
$A_{\rm  radio}$  to  a  radio  activity class  (HRA  or  LRA).   Upon
inspection, these fractions are qualitatively consistent with what one
would  expect if  radio and  optical activity  are  independent, i.e.,
$f(A_{\rm opt}+A_{\rm radio}|M_{\rm h},{\rm C}) = f(A_{\rm opt}|M_{\rm
  h},{\rm C})  \times f(A_{\rm radio}|M_{\rm h},{\rm  C})$.  Thus, for
example,  central radio  galaxies  with star  formation are  typically
found in less massive haloes  than central radio galaxies with optical
AGN activity, irrespective of their specific radio power.  In the case
of satellite galaxies, the statistics  are typically too poor to reach
any meaningful  conclusion.  This finding  that the optical  and radio
activities of galaxies are independent  is in good agreement with Best
et al. (2005a), who found that  the probability that a galaxy of given
stellar  mass  classifies as  HRA  is  independent  of whether  it  it
optically classified as an AGN.

\subsection{Stellar Mass Dependence}
\label{sec:stellarmass}

\begin{figure*}
\centerline{\psfig{figure=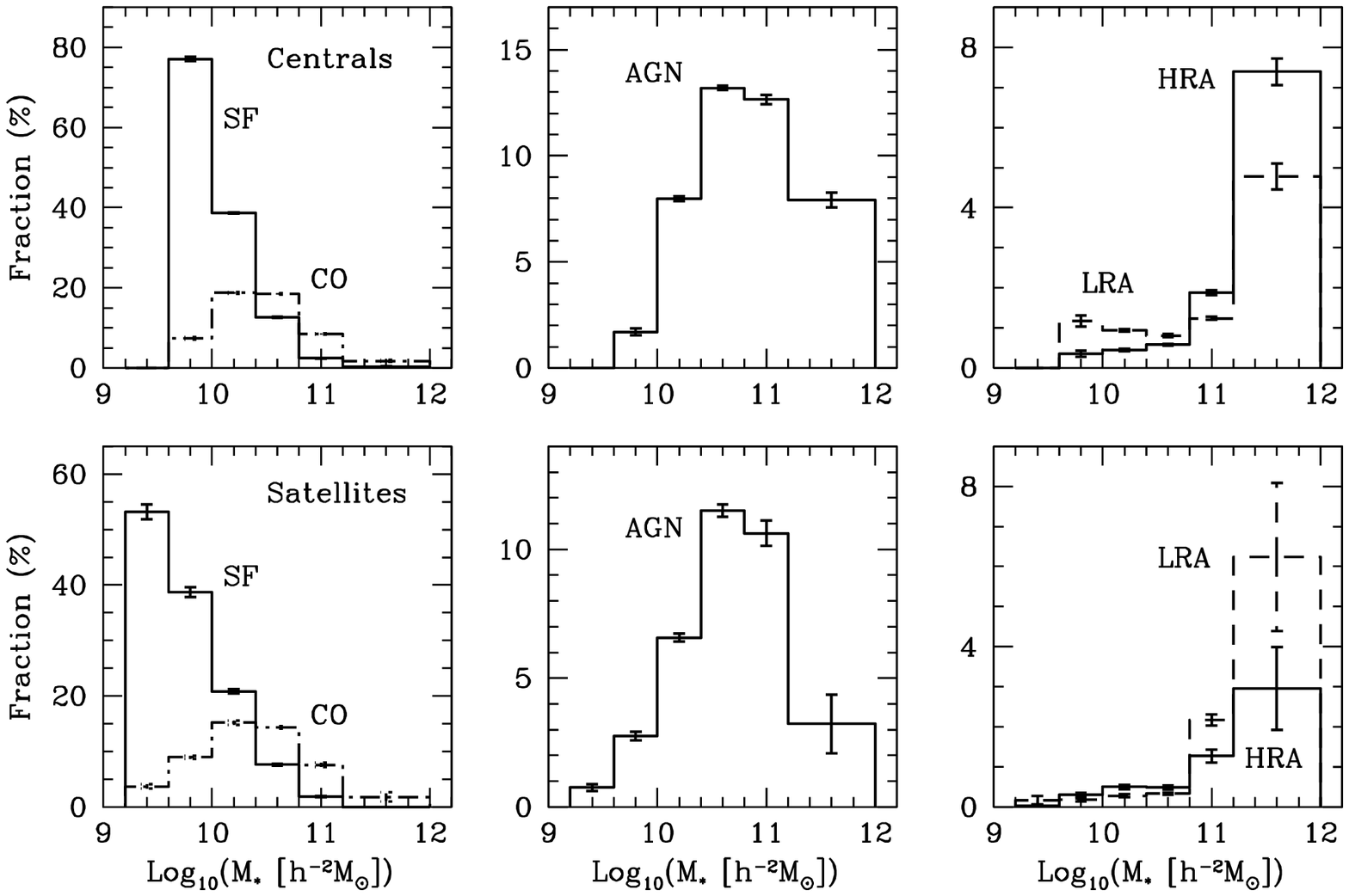,width=16cm}}
\caption{The activity fractions $f(A|M_{\ast},{\rm C})$ for centrals 
  (upper  panels) and $f(A|M_{\ast},{\rm  S})$ for  satellite galaxies
  (lower  panels), as  functions of  their stellar  mass,  $M_{\ast}$. 
  These  fractions  reflect  the  stellar mass  distributions  of  the
  various activity classes (SF, CO, AGN, LRA and HRA, as labeled).}
\label{fig:stellarmass}
\end{figure*}

We  now turn our  attention to  the {\it  stellar} mass  dependence of
galaxy    activity.      The    upper    and     lower    panels    of
Fig.~\ref{fig:stellarmass}    show    $f(A|M_{\ast},{\rm   C})$    and
$f(A|M_{\ast},{\rm S})$, respectively, where $M_{\ast}$ is the stellar
mass  of the  galaxy.  Note  how the  typical stellar  mass  of active
centrals increases  going from star forming centrals  to centrals with
optical  AGN  activity to  radio-active  centrals.   The stellar  mass
distribution of SF centrals peaks at $\sim 10^{9.8} h^{-2} \Msun$, and
drops  rapidly  with increasing  $M_{\ast}$,  becoming negligible  for
$M_{\ast} \gta 10^{11} h^{-2}  \Msun$.  The stellar mass distributions
of CO and AGN centrals  peak near $10^{10.6} h^{-2} \Msun$, though the
distribution  of the  AGN  centrals is  somewhat  more skewed  towards
larger  masses.  Finally,  central  galaxies with  radio activity  are
typically very massive, with stellar masses in excess of $\sim 10^{11}
h^{-2} \Msun$.  Qualitatively, these  trends are remarkably similar to
those   with  halo  mass   shown  in   Fig.~\ref{fig:halomass}.   This
similarity owes  to the fact  that the stellar  mass and halo  mass of
central galaxies are tightly related  (e.g., Yang, Mo \& van den Bosch
2008), and indicates that it  will be difficult to disentangle whether
the activity of central galaxies is causally connected to halo mass or
to stellar mass (see below).

As for  the satellites, the stellar mass  distributions are remarkably
similar to those of the centrals. This is very different from the halo
mass  distributions (cf.  Fig.~\ref{fig:halomass}),  which show  clear
differences for  centrals and satellites.  This is  a first indication
that the activity of a galaxy  is more closely related to stellar mass
than to halo mass.

Finally,  we note  that  the stellar  mass  distributions for  various
activity  classes  shown  in  Fig.~\ref{fig:stellarmass} are  in  good
agreement  with  previous  studies  (e.g., Brinchmann  et  al.   2004;
Kauffmann et al. 2004; Best et al. 2005b; Kewley et al. 2006).

\subsection{Breaking the causality degeneracy}
\label{sec:degeneracy}

\begin{figure*}
\centerline{\psfig{figure=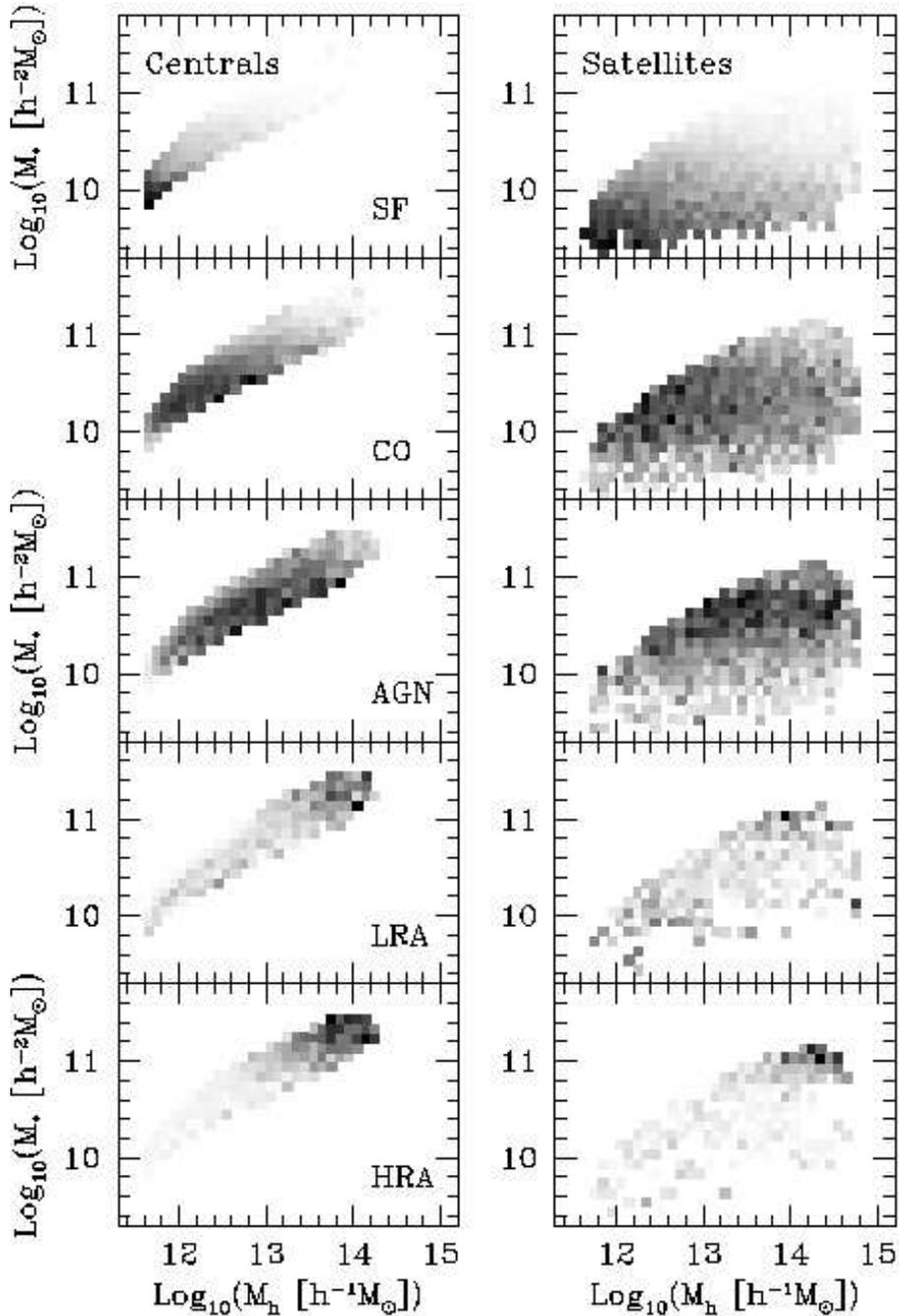,width=12cm}}
\caption{The activity fractions $f(A|M_{\rm h},M_{\ast},C)$ (left-hand 
  panels)  and  $f(A|M_{\rm  h},M_{\ast},S)$  (right-hand  panels)  as
  functions of halo mass and stellar  mass. Only bins with at least 50
  galaxies  are shown, with  darker shades  of grey  indicating larger
  fractions.  From top to bottom,  results are shown for the following
  activity  classes: SF,  CO,  AGN,  LRA and  HRA,  as indicated.  For
  central galaxies,  the stellar mass is tightly  correlated with halo
  mass, preventing us from investigating which of these two parameters
  is  more causally  connected  to the  activity.   For the  satellite
  galaxies, however, the fractions are almost independent of halo mass
  at fixed stellar mass, indicating that stellar mass is more causally
  connected to galaxy activity than halo mass.}
\label{fig:degeneracy}
\end{figure*}

In order to examine whether halo mass or stellar mass is more causally
connected  to  galaxy activity,  one  needs  to  probe the  halo  mass
dependence  for   fixed  bins  in   stellar  mass  and  vice   versa.  
Fig.~\ref{fig:degeneracy}    plots     the    fractions    $f(A|M_{\rm
  h},M_{\ast},C)$  (left-hand panels) and  $f(A|M_{\rm h},M_{\ast},S)$
(right-hand panels) as functions of  halo mass and stellar mass.  Only
$(M_{\rm h},M_{\ast})$-bins with at  least 50 galaxies are shown, with
darker  shades of grey  indicating larger  fractions.  Note  that this
figure reveals the same  overall trends as in Figs.~\ref{fig:halomass}
and~\ref{fig:stellarmass},      which      are     projections      of
Fig.~\ref{fig:degeneracy} along  the $M_{\ast}$ and  $M_{\rm h}$ axes,
respectively.

\begin{figure*}
\centerline{\psfig{figure=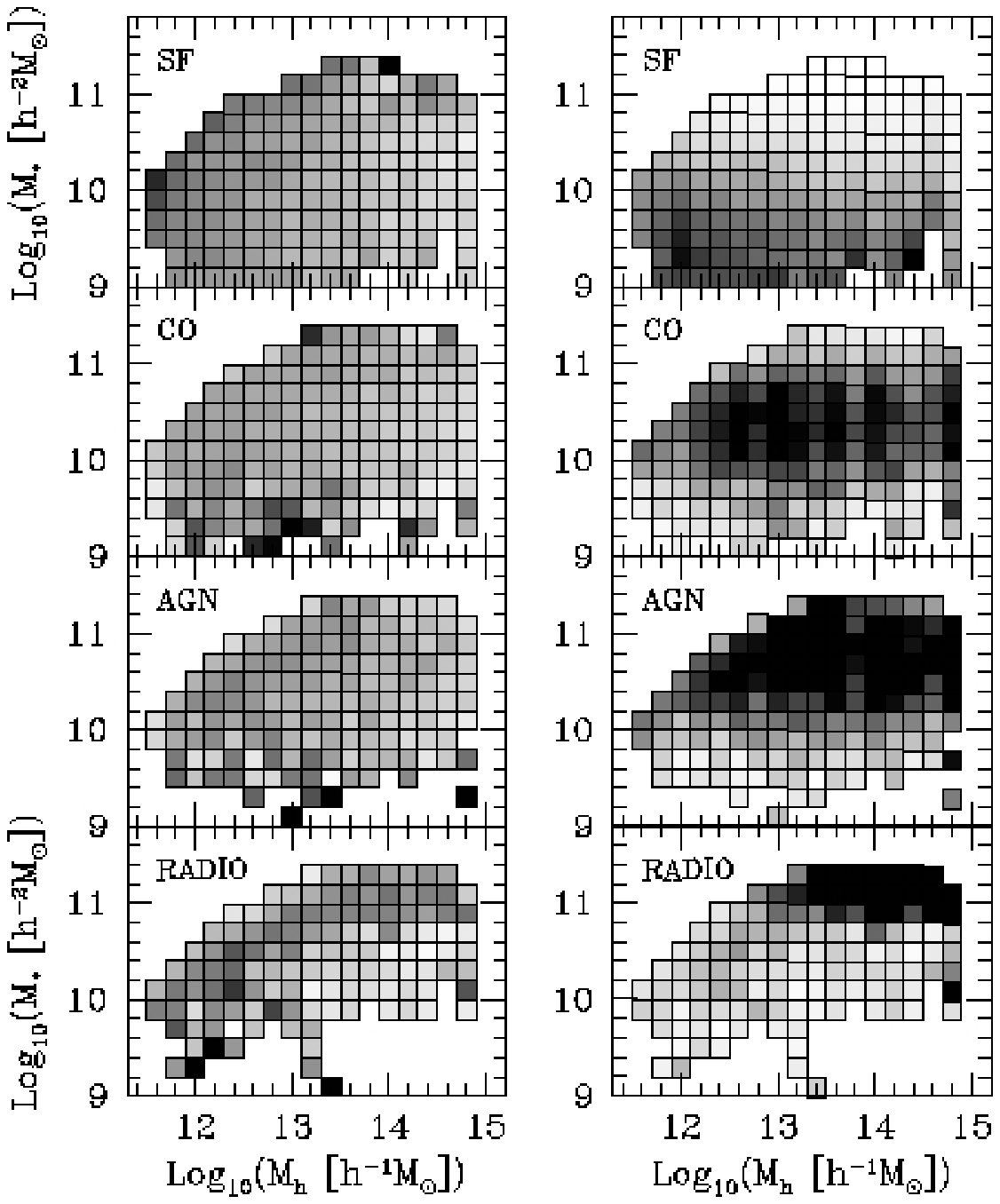,width=10cm}}
\caption{The reduced conditional probability functions, $Q_1$
  (left-hand panels)  and $Q_2$  (right-hand panels), as  functions of
  both  stellar mass  and halo  masses.  Only  bins with  at  least 50
  galaxies  are shown, with  darker shades  of grey  indicating larger
  values of $Q$.  No distinction is made between central and satellite
  galaxies.  From top  to bottom, results are shown  for the following
  activity classes: SF,  CO, AGN and RADIO (LRA  and HRA combined), as
  indicated.  By  definition, any variation of $Q_1$  with $M_{\rm h}$
  at  fixed $M_{\ast}$ indicates  a dependence  of galaxy  activity on
  halo  mass.  Conversely,  a variation  of $Q_2$  with  $M_{\ast}$ at
  fixed $M_{\rm h}$ indicates a  dependence of galaxy activity on halo
  mass. It is  clear from this plot that stellar  mass is more tightly
  related to galaxy activity than halo mass.}
\label{fig:QoneQtwo}
\end{figure*}

The relatively  tight correlation between  stellar mass and  halo mass
for centrals, which partly owes to  the way that halo masses have been
assigned to  the groups  (see Section~\ref{sec:data} and  Y07), limits
the dynamic range over which we  can probe the halo mass dependence at
fixed stellar mass, and vice  versa. Hence, it is virtually impossible
to tell which of these  two conditionals is more causally connected to
the activity. In the case of the satellites, however, it is clear that
the fractions vary more with stellar mass at fixed halo mass than with
halo mass  at fixed stellar  mass, irrespective of the  activity class
being  considered. In fact,  the fractions  are almost  independent of
halo mass at fixed stellar  mass, which clearly indicates that stellar
mass is  a more defining property  of the activity class  to which the
galaxy belongs than halo mass.

To make  this more quantitatively,  we ignore the  distinction between
centrals and satellites and define the reduced conditional fractions
\begin{equation}
Q_1 = \frac{f(A|M_{\ast},M_{\rm h})}{f(A|M_*)} \,,
\end{equation}
and 
\begin{equation}
Q_2 = \frac{f(A|M_{\rm h},M_{\ast})}{f(A|M_{\rm h})}\,.
\end{equation}
Note that any variation of $Q_1$ as a function of $M_{\rm h}$ at fixed
$M_{\ast}$  traces the  dependence of  galaxy activity  on  halo mass;
conversely, changes in  $Q_2$ as a function of  $M_*$ at fixed $M_{\rm
  h}$  reflect that  galaxy  activity depends  on  stellar mass.   The
reduced fractions  $Q_1$ and  $Q_2$ are plotted  in the  left-hand and
right-hand  panels  of  Fig.~\ref{fig:QoneQtwo},  respectively,  as  a
function  of halo  mass and  stellar mass  for the  different activity
classes.  Note  that we have combined  the HRA and LRA  classed into a
single  class (labeled  RADIO) in  order to  increase  the statistical
significance of the present analysis. As in Fig.~\ref{fig:degeneracy},
darker shades of grey correspond to larger values of $Q_1$ and $Q_2$.

\begin{figure*}
\centerline{\psfig{figure=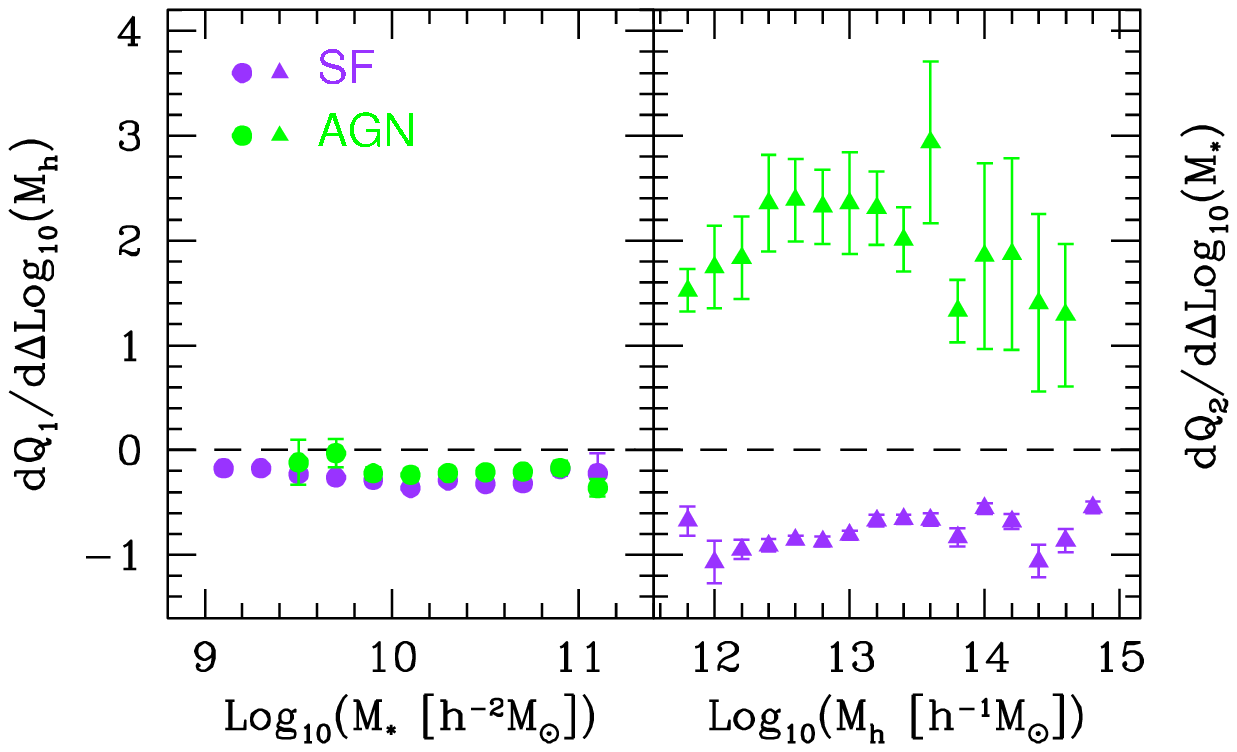,width=10cm}}
\caption{The slopes ${\rm  d} Q_1/ {\rm d}\log  M_{\rm h}$ as function
  of  $M_{\ast}$  (left-hand panel)  and  ${\rm  d}  Q_2/ {\rm  d}\log
  M_{\ast}$  as function  of  $M_{\rm h}$  (right-hand  panel) for  SF
  galaxies  (purple symbols)  and  optical AGN  (green symbols).   The
  horizontal dashed lines correspond to  slopes of zero, and are shown
  for comparison. Note  that $\vert {\rm d} Q_2/  {\rm d}\log M_{\ast}
  \vert$ is significantly larger than ${\rm d} Q_1/ {\rm d}\log M_{\rm
    h}$ emphasizing  that the  occurrence of SF  and AGN  activity are
  mainly  related  to galaxy  stellar  mass,  rather  than the  galaxy
  environment.}
\label{fig:Qslope}
\end{figure*}

While at  a given  stellar mass, $Q_1$  varies little with  halo mass,
$Q_2$ changes dramatically  with stellar mass at a  given $M_{\rm h}$. 
This  confirms that  galaxy  activity is  more  causally connected  to
stellar mass than to halo mass.  For the star forming galaxies and the
optical AGN, we find that  the $Q_2(\log M_{\ast}|\log M_{\rm h})$ are
well  fitted by  simple  linear relations.   The  right-hand panel  of
Fig.~\ref{fig:Qslope} plots the corresponding slopes, ${\rm d}Q_2/{\rm
  d}\log M_{\ast}$, as functions of $M_{\rm h}$. Clearly, these differ
significantly  from  zero,  with   little  dependence  on  halo  mass,
indicating  that the  fraction of  galaxies  with SF  and optical  AGN
activity depends  strongly on stellar  mass at fixed $M_{\rm  h}$. For
comparison, the left-hand panels  plots ${\rm d}Q_1/{\rm d}\log M_{\rm
  h}$  as functions  of  $M_{\ast}$  for the  same  activity classes.  
Although significantly different from zero (dashed line), these slopes
are much smaller.  The average ${\rm d} Q_1/ {\rm d}\log M_{\rm h}$ is
$-0.25$  and  $-0.20$  for  the  SF and  AGN  galaxies,  respectively,
indicating  that there  is a  weak trend  for SF  and AGN  activity to
decrease with  increasing halo  mass. On the  other hand,  the average
values of  ${\rm d} Q_2/ {\rm  d}\log M_{\ast}$ are  $-0.80$ and $2.0$
for  SF  and AGN  galaxies,  respectively,  indicating  that while  SF
activity decreases with increasing  stellar mass, optical AGN activity
is strongly boosted in more massive galaxies.  Comparing these values,
the stellar mass dependencies of SF and optical AGN activity are $\sim
3$ and $\sim 10$ times more prominent than the corresponding halo mass
dependencies.   This   is  the  quantitative   confirmation  that  the
occurrence  of galaxy  activity  is more  strongly  related to  galaxy
stellar mass than to halo mass.

\begin{figure}
\centerline{\psfig{figure=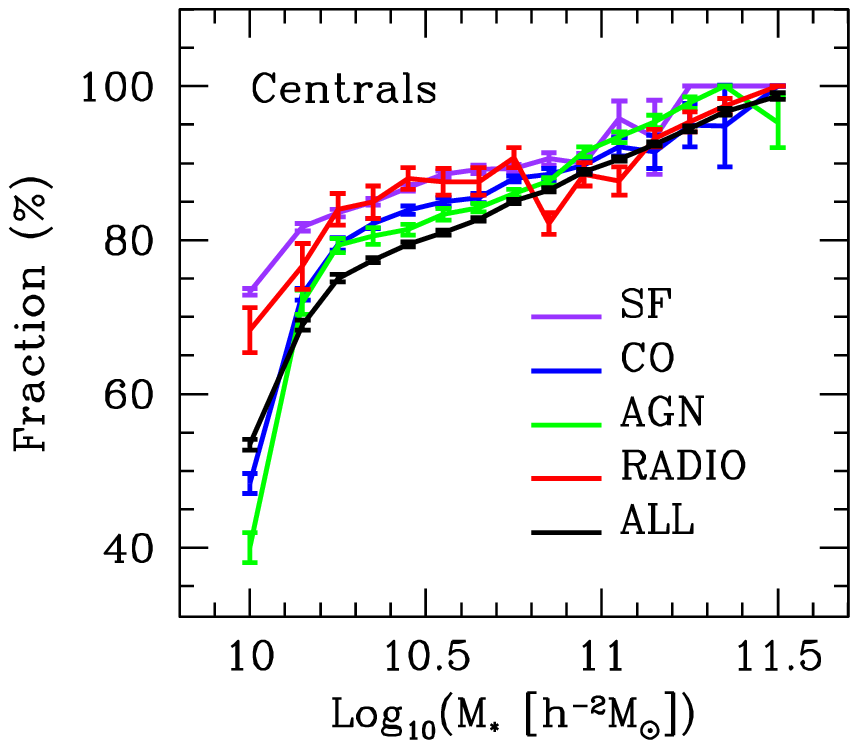,width=8cm}}
\caption{The central fractions, $f_{\rm cen}$ as functions of stellar
  mass for all  galaxies (black line), and for  four activity classes,
  as indicated. Note that being active increases the probability for a
  galaxy to be a central,  indicating that being located at the center
  of the halo boosts both the star formation and AGN activity.}
\label{fig:censat}
\end{figure}

\subsection{Group hierarchy}
\label{sec:hierarchy}

Our analysis  thus far has clearly  shown that stellar mass  is a more
important parameter  for determining the  activity class of  a certain
galaxy than halo mass. We  now investigate whether the group hierarchy
plays a  role; does the mere fact  that a galaxy is  a central galaxy,
rather than a  satellite galaxy, boost or suppress  its probability to
be active in the optical and/or radio? 

To address this question we have computed the central fractions
\begin{equation}
f_{\rm cen}(A,M_{\ast}) = \sum\limits_{i=1}^{N_{\rm cen}} w_i /
         \sum\limits_{i=1}^{N_{\rm cen}+N_{\rm sat}} w_i\,,
\end{equation}
with  $N_{\rm cen}$  and  $N_{\rm  sat}$ the  numbers  of central  and
satellite galaxies of activity class  $A$ and stellar mass $M_{\ast}$. 
Once again, the  errors on the central fractions  are determined using
the jackknife  technique with 20  subsamples. The results for  SF, CO,
AGN   and    RADIO   (HRA   and    LRA   combined)   are    shown   in
Fig.~\ref{fig:censat} as indicated.  For  comparison, we also show the
central fractions of {\it all} galaxies, independent of their activity
(or  absence  thereof).   Clearly,  the central  fractions  of  active
galaxies  are higher  than those  of  all galaxies,  at basically  all
stellar  masses, and for  all activity  classes.  This  indicates that
galaxies of  a given stellar  mass are more  likely to be  active when
they are centrals than when  they are satellites.  This effect is most
pronounced for star formation, while  optical AGN activity seems to be
only  boosted  by a  mild  amount  for  central galaxies  compared  to
satellites of the same stellar mass.

This is in  qualitative agreement with Best et  al.  (2007), who found
that brightest  cluster galaxies  (i.e. equivalent to  our `centrals')
are more  likely to host a  radio-loud AGN than other  galaxies of the
same  stellar mass, and  with van  den Bosch  \etal (2008a),  who have
shown  that central  galaxies  have bluer  colors,  on average,  than
satellite galaxies of the  same stellar mass. Since satellite galaxies
of a  given stellar  mass where central  galaxies of the  same stellar
mass prior  to being accreted by  their host halo,  this suggests that
activity, whether  in the optical or  in the radio,  is suppressed (or
quenched) as  soon as a galaxy  becomes a satellite.

\begin{figure*}
\centerline{\psfig{figure=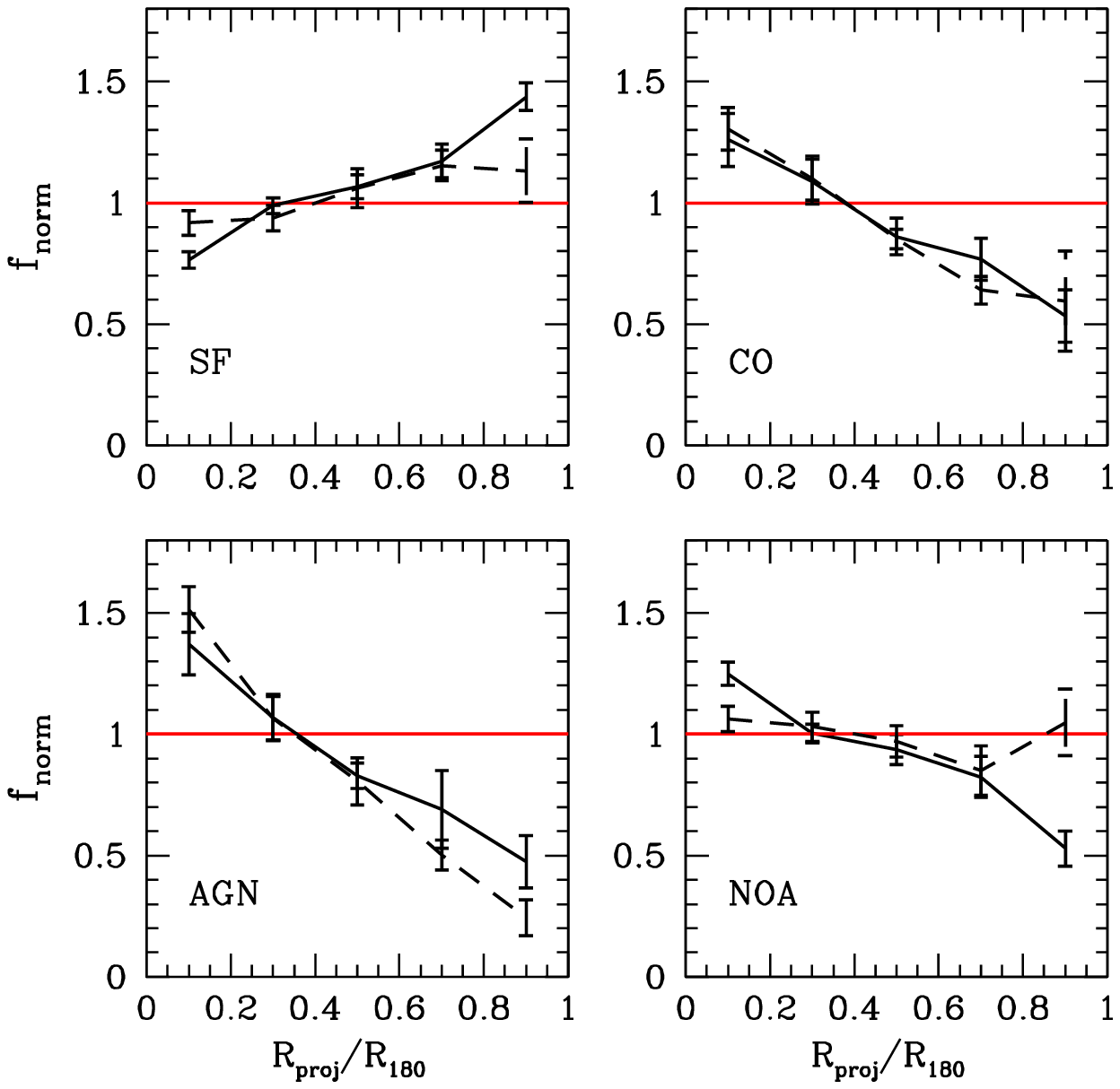,width=10cm}}
\caption{The solid lines show the radial distributions of satellite
  galaxies  of different  activity classes,  corrected for  the radial
  distribution of  the general population  of satellites unconstrained
  in activity (black solid line).   Results are shown for star forming
  satellites (SF), composite  satellites (CO), satellites with optical
  AGN activity (AGN),  and satellites with no optical  activity (NOA).
  If active  satellites follow the  same radial distribution  as their
  parent  population, they  would follow  the red  horizontal  line at
  $f_{\rm norm}=1$.  Clearly, CO and AGN satellites are more centrally
  concentrated than their parent  population, while SF satellites have
  a preference  to populate the  outskirts of their parent  haloes The
  dashed lines indicate the  radial distribution of satellite galaxies
  that  have  the same  stellar  masses  as  those in  the  respective
  activity classes. The fact that  these are very similar to the solid
  lines  indicates   that  activity  is  not   causally  connected  to
  halo-centric  distance; rather  the  trends shown  are  due to  mass
  segregation. See text for a detailed discussion.}
\label{fig:radius}
\end{figure*}

\subsection{Halo-centric Distance}
\label{sec:radius}

The  galaxy  group catalogue  also  allows  us  to check  whether  the
position of a galaxy within its  group may foster or hamper a specific
kind of activity. For this purpose, we compute the projected distance,
$R_{\rm proj}$, of each  satellite galaxy from the luminosity-weighted
center of  its group  normalized by the  characteristic radius  of the
group,   $R_{180}$  (see   Section~\ref{sec:data}).   We   define  the
normalized fraction  of satellites of  a given activity class  and per
bin of $R_{\rm norm} \equiv R_{\rm proj}/R_{180}$ as:
\begin{equation}\label{radiusnorm}
f_{\rm norm} = \frac{f(R_{\rm norm}|A,{\rm S})}{f(R_{\rm norm}|{\rm S})}
\end{equation}
where  $f(R_{\rm  norm}|A,{\rm  S})$   is  the  weighted  fraction  of
satellite galaxies  of activity  class $A$ in  a given bin  of $R_{\rm
  norm}$,  and  $f(R_{\rm  norm}|S)$   is  the  weighted  fraction  of
satellite  galaxies (unconstrained  by activity)  in the  same $R_{\rm
  norm}$-bin. This  distribution is therefore  normalized with respect
to  the radial  distribution of  the general  population  of satellite
galaxies.

The  solid  curves in  Fig.~\ref{fig:radius}  show  $f_{\rm norm}$  as
function  of  $R_{\rm  norm}$,   thus  obtained.   Results  are  shown
separately for  star forming  satellites (SF; upper  left-hand panel),
composite  satellites (CO;  upper right-hand  panel),  satellites with
optical AGN activity (AGN;  lower left-hand panel), and satellite with
no  optical activity (NOA;  lower right-hand  panel).  No  results are
shown  for radio  activity,  simply because  the  number of  satellite
galaxies with  radio activity is too  small to allow  for a meaningful
assessment of  their radial  distribution. As before,  the (1$\sigma$)
errors have been computed using the jackknife method.

The  SF satellites  have $f_{\rm  norm}<1$  ($> 1$)  at small  (large)
$R_{\rm norm}$.   This suggests that  star formation is  suppressed at
small halo-centric  radii.  CO  and AGN activity,  on the  other hand,
seems  to  be boosted  at  small  halo-centric  radii. Although  these
results seem  to indicate that halo-centric radius  plays an important
causal role  for galaxy activity, it  is important to  realize that we
have not used stellar mass as  a control variable. As shown in van den
Bosch et  al.  (2008b), the group  catalogue of Y07  used here reveals
clear evidence  for mass segregation, in that  more massive satellites
preferentially  reside at smaller  halo-centric distances.   Since the
analysis  in Section~\ref{sec:degeneracy}  indicates that  there  is a
strong causal connection between stellar mass and galaxy activity, the
trends with halo-centric radius  shown in Fig.~\ref{fig:radius} do not
necessarily  imply  that activity  is  causally  connected to  $R_{\rm
  norm}$.   In order to  test this  we proceed  as follows.   For each
satellite  galaxy of  activity class  $A$ in  a given  bin  of $R_{\rm
  norm}$ we  select a random satellite, unconstrained in activity, in
the same  radial bin (but not  necessarily in the same  group) that is
matched in stellar mass ($\Delta\log M_{\ast} \leq 0.05$) and redshift
($\Delta z \leq 0.01$).  We use this matched sample to compute
\begin{equation}
f'_{\rm norm} = \frac{f_{M}(R_{\rm norm}|{\rm S})}{f(R_{\rm norm}|{\rm S})}
\end{equation}
where  $f_{M}(R_{\rm  norm}|{\rm S})$  is  the  fraction of  satellite
galaxies (unconstrained in  activity) with halo-centric radius $R_{\rm
  norm}$ that  are matched  in stellar mass  and redshift to  those in
$f(R_{\rm norm}|A,{\rm  S})$. The results  thus obtained are  shown in
Fig.~\ref{fig:radius}  as dashed  lines.   Overall, it  is clear  that
$f'_{\rm norm}(R_{\rm norm})$ is  very similar to $f_{\rm norm}(R_{\rm
  norm})$,  for all activity  classes shown.  This indicates  that the
trends with halo-centric radius of  the latter are simply a reflection
of  the  mass segregation  combined  with  a  causal relation  between
activity and  stellar mass.   In other words,  there is  no indication
that the occurrence of galaxy activity {\it in satellite galaxies} has
a direct dependence on halo-centric radius.

\begin{figure*}
\centerline{\psfig{figure=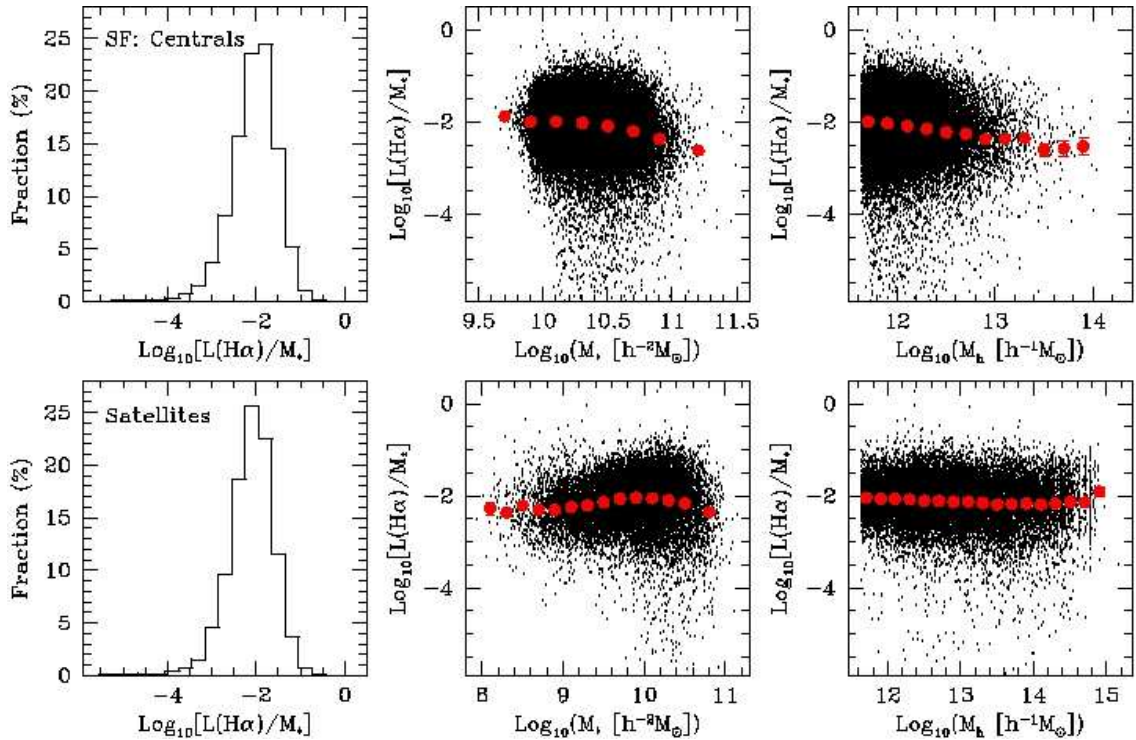,width=15cm}}
\caption{{\it Left-hand panels:} the distributions of
  $\log[L({\rm  H}\alpha)/M_{\ast}]$  for  star-forming  centrals  and
  satellites, normalized by their  total number.  {\it Middle panels:}
  scatter plots of the specific  H$\alpha$ luminosity as a function of
  stellar  mass  for  centrals  (upper panel)  and  satellites  (lower
  panel).   The  solid  red  circles indicate  the  mean  $\log[L({\rm
    H}\alpha)/M_{\ast}]$ per  bin of stellar mass,  while the errorbar
  indicates the error  on the mean.  {\it Right-hand  panels:} same as
  in the middle panels, but as function of halo mass, $M_{\rm h}$.}
\label{fig:SFstrength}
\end{figure*}

\section{The Ecology of Galaxy Activity Strength}
\label{sec:strength}

In the previous section we have  shown that the occurrence of any kind
of galaxy  activity is  mainly governed by  stellar mass,  rather than
halo mass.  In  addition, there is an indication  that being a central
galaxy, rather  than a  satellite of the  same stellar mass,  causes a
significant boost in the occurrence of activity.

We now investigate  the dependence of the  activity {\it strength}  on
stellar mass, halo mass and  group  hierarchy. For star formation  and
optical AGN activity, we define the activity  strength in terms of the
specific luminosities in  the   H$\alpha$ and [OIII]   emission lines,
$L({\rm   H}\alpha)/M_{\ast}$      and   $L[{\rm     OIII}]/M_{\ast}$,
respectively, both   corrected  for dust obscuration   as described in
Section 2. Since the spectra are  obtained with fibers with a diameter
of 3$''$, we need  to correct $L({\rm  H}\alpha)$ for the missed flux.
To that extent  we scale $L({\rm  H}\alpha)$ by the ratio  measured in
the $r$ band  between the galaxy Petrosian  flux and the flux observed
through  the spectroscopic  fiber.   Note that  no such  correction is
needed for  $L[{\rm OIII}]$, since  the [OIII] emission is believed to
be dominated by emission from the AGN.

The results for the  SF strength $L({\rm H}\alpha)/M_{\ast}$ are shown
in Fig.~\ref{fig:SFstrength}.  The  histograms in the left-hand panels
show the  distributions of $\log[L({\rm H}\alpha)/M_\ast]$  for the SF
centrals (upper  left-hand panel)  and SF satellites  (lower left-hand
panel), normalized by the total  numbers of centrals and satellites in
our  group  sample,  respectively.   Both  distributions  peak  around
$L({\rm H}\alpha)/M_\ast \sim 10^{-2} L_{\odot}/(h^{-2}\Msun)$ and are
weakly  skewed  towards   smaller  specific  luminosities.   No  clear
differences  are evident  between the  distributions for  centrals and
satellites.      The    middle     and     right-hand    panels     of
Fig.~\ref{fig:SFstrength} show scatter plots of the specific H$\alpha$
luminosities as functions of stellar mass and halo mass, respectively.
The  solid   red  circles   indicate  the  averages   of  $\log[L({\rm
  H}\alpha)/M_\ast]$ per bin  of stellar mass or halo  mass, while the
errorbars  (typically smaller  than  the solid  circles) indicate  the
error on  the mean.   For central galaxies,  the mean  of $\log[L({\rm
  H}\alpha)/M_\ast]$  decreases by  a factor  of four  with increasing
stellar mass and increasing halo mass.  These trends are statistically
significant at $\sim$ 4$\sigma$.  Thus, in addition to suppressing the
{\it occurrence}  of SF  activity, a larger  stellar mass or  a larger
host halo mass also suppresses the  {\it strength} of the SF activity. 
Unfortunately, the sample is not  large enough to examine whether halo
mass  or stellar mass  are more  causally linked  to this  trend.  For
satellite galaxies, we find no obvious trend of the specific H$\alpha$
luminosity with  halo mass or  stellar mass.  This  difference between
centrals  and satellites suggests  that group  hierarchy has  a direct
impact  on  the strength  of  star  formation  activity in  galaxies.  

\begin{figure*}
\centerline{\psfig{figure=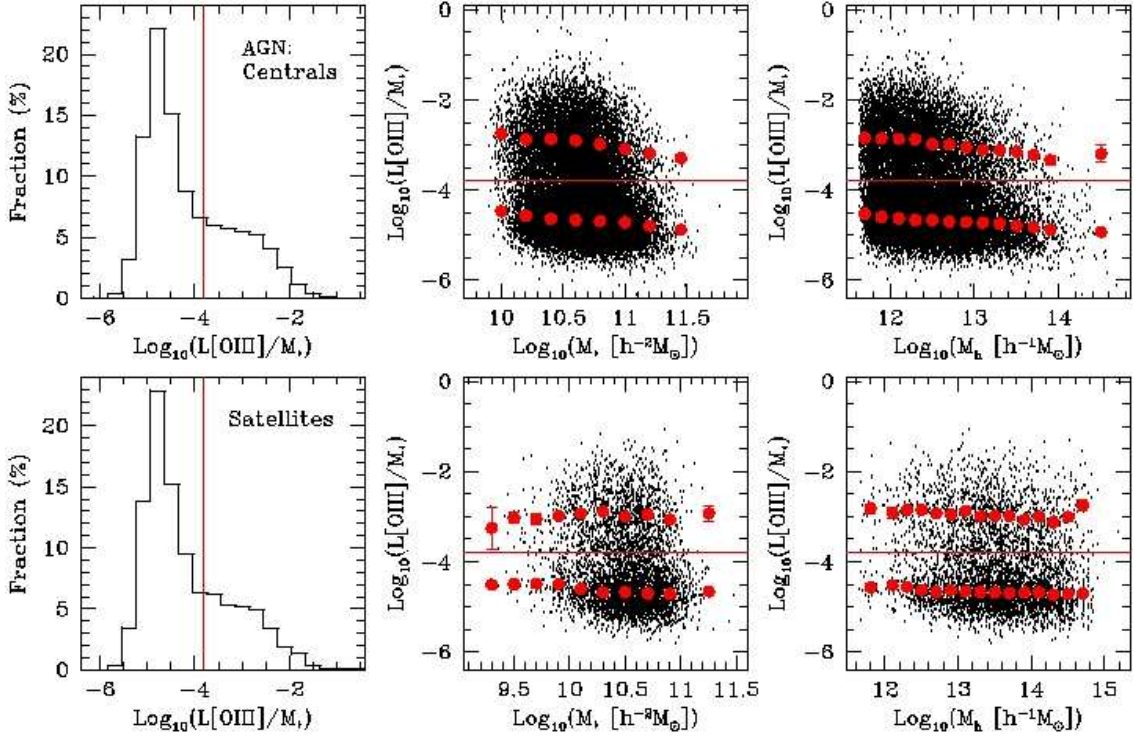,width=15cm}}
\caption{Same as Fig.~\ref{fig:SFstrength} but for the specific
  [OIII]  luminosity of  optical AGN.   The vertical  (horizontal) red
  line in the left-hand  (middle and right-hand) panels corresponds to
  $\log(L[{\rm OIII}]/M_{\ast})  = -3.8$,  and indicates our  split in
  LINERS  ($\log(L[{\rm  OIII}]/M_{\ast})  \leq  -3.8$)  and  type  II
  Seyferts ($\log(L[{\rm OIII}]/M_{\ast}) > -3.8$).}
\label{fig:AGNstrength}
\end{figure*}

\begin{figure*}
\centerline{\psfig{figure=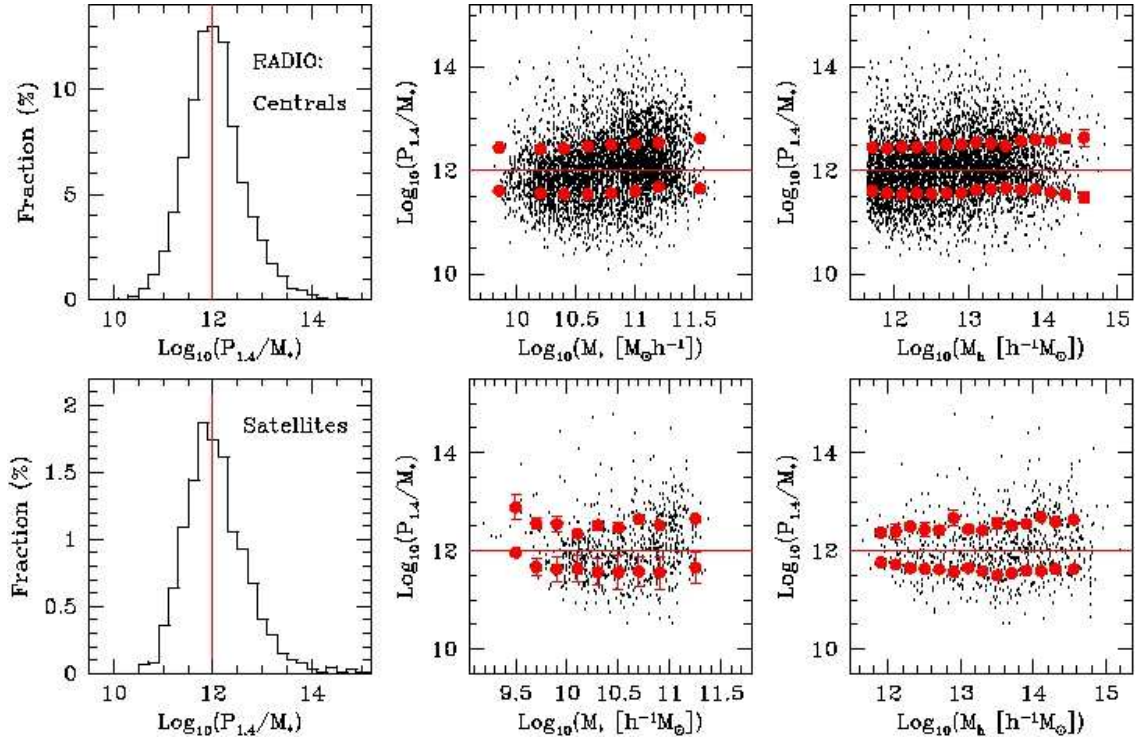,width=15cm}}
\caption{Same as Fig.~\ref{fig:SFstrength} but for the specific
  radio power  of radio galaxies.  The vertical  (horizontal) red line
  in  the  left-hand (middle  and  right-hand)  panels corresponds  to
  $\log(P_{1.4}/M_{\ast}) = 12$, and indicates our split in high radio
  activity (HRA;  $\log(L[{\rm OIII}]/M_{\ast})  > 12$) and  low radio
  activity (LRA; $\log(L[{\rm OIII}]/M_{\ast}) \leq 12$) galaxies.}
\label{fig:RADIOstrength}
\end{figure*}

Fig.~\ref{fig:AGNstrength} shows the same as Fig.~\ref{fig:SFstrength}
but  for  the  specific  [OIII]  luminosity (related  to  optical  AGN
activity) rather  than the  specific H$\alpha$ luminosity.   Note that
the  distributions  of  $\log(L[{\rm  OIII}]/M_{\ast})$  are  strongly
skewed  towards higher values,  for both  centrals and  satellites. In
fact, the distributions  are well fit by a sum  of two Gaussians, with
their respective peaks at $\log(L[{\rm OIII}]/M_{\ast}) \simeq -5$ and
$-3$.  We  identify these  two `populations' with  LINERS and  type II
Seyfert galaxies, respectively (cf. Kewley et al.  2006).  In order to
probe  these  two populations  separately,  we (somewhat  arbitrarily)
split our optical AGN in LINERS and type II Seyfert galaxies depending
on whether  $\log(L[{\rm OIII}]/M_{\ast})$  is smaller or  larger than
$-3.8$,  respectively. As  is evident  from  the upper  panels in  the
middle  and  right-hand  columns  of  Fig.~\ref{fig:AGNstrength},  the
central galaxies with  Seyfert II activity reveal a  decrease of their
average specific  [OIII] luminosity by  a factor three going  from low
mass  (either  stellar  or  halo)  to  high  mass  (at  the  4$\sigma$
significance level).   For the central LINERS,  the (average) specific
[OIII] luminosity  diminishes by  a factor of  about 2  for increasing
stellar and halo  mass (at the 6$\sigma$ significance  level).  On the
contrary,  satellite  galaxies  show  no  dependence  of  optical  AGN
strength  on either stellar  mass or  halo mass;  neither for  type II
Seyferts nor for LINERS.  Finally,
Fig.~\ref{fig:RADIOstrength}       shows       the       same       as
Figs.~\ref{fig:SFstrength}   and~\ref{fig:AGNstrength}  but   for  the
specific        radio        power       $P_{1.4}/M_{\ast}$        [in
W~Hz$^{-1}/$($h^{-2}$M$_{\odot}$)].   As  evident  from the  left-hand
panels, the  distributions of specific  radio power for  both centrals
and satellites  peak at $\log(P_{1.4}/M_{\ast}) = 12$  (indicated by a
red,  vertical  line),  and we  have  used  this  value to  split  the
population   of  radio   galaxies   in  high   radio  activity   (HRA;
$\log[P_{1.4}/M_{\ast}]   >  12$)   and  low   radio   activity  (LRA:
$\log[P_{1.4}/M_{\ast}] \leq 12$) subsamples.
The mean  specific radio power  of HRA centrals increases  mildly with
stellar mass by a factor of 1.5 (at the $3\sigma$ significance level),
while the mean  specific radio power of LRA  centrals is constant with
$M_{\ast}$.  In the  case of radio satellites (both  HRA and LRA), the
mean   $\log(P_{1.4}/M_{\ast})$,  although   noisy,  shows   no  clear
dependence on  stellar mass.  Similar  trends are seen as  function of
halo mass (right-hand panels of Fig.~\ref{fig:RADIOstrength}).

\section{Conclusions}
\label{sec:concl}

An important aspect of galaxy formation is understanding the processes
that  trigger, quench  and control  star formation  and AGN  activity. 
Since  both  require   the  presence  of  gas,  this   is  similar  to
understanding  how galaxies  accrete their  gas,  and how  gas can  be
expelled,  stripped or  exhausted.  A  large number  of  processes are
thought  to play a  role. At  the low  mass end,  it is  believed that
supernova feedback  plays a crucial  role in expelling  and/or heating
gas so  as to keep the  overall star formation efficiency  low. At the
massive end, it is generally believed that AGN activity somehow expels
or  reheats cold  gas  causing  a quenching  of  star formation.   Ram
pressure and  tidal forces are  believed to remove gas  from satellite
galaxies, while  major mergers are  efficient in funneling gas  to the
central  regions, igniting  starbursts and/or  AGN activity,  which in
turn  may exhaust  and/or  expel  the gas  resulting  in a  subsequent
shutdown of  galaxy activity.  In  order to probe the  environments in
which these  various processes operate,  we have used the  SDSS galaxy
group catalogue of Yang et  al.  (2007) and examined the dependence of
galaxy  activity  on stellar  mass,  halo  mass,  and group  hierarchy
(centrals vs.   satellites). In particular,  we have split  the galaxy
population in star formers, optical AGN, composite galaxies (both star
formation and optical  AGN activity) and radio sources  (split in high
and low radio activity) using  emission line fluxes extracted from the
SDSS spectra by Kauffmann et al.  (2003b) and by cross correlating the
SDSS with the FIRST and NVSS catalogues.

For the subsample of central galaxies, we  find a smooth transition in
halo mass  as the   activity of  central  galaxies  changes  from star
formation  to   optical    AGN  activity  to   radio    emission  (see
Fig.~\ref{fig:halomass}).  Star-forming centrals preferentially reside
in  haloes less massive  than  $10^{12} h^{-1}\Msun$, central galaxies
with optical AGN activity inhabit haloes with a characteristic mass of
$\sim  10^{13}  h^{-1} \Msun$  and   centrals emitting  in  the  radio
preferentially reside  in haloes with $M_{\rm   h} \gta 10^{14} h^{-1}
\Msun$. A similar trend is present as function  of stellar mass.  Star
formation is mainly occurring in  central galaxies with $M_{\ast} \lta
10^{10}   h^{-2} \Msun$,  while the  hosts   of  optical-AGN and radio
sources have   typical stellar masses   of $10^{10.8} h^{-2}\Msun$ and
$10^{11.6}            h^{-2}\Msun$,       respectively            (see
Fig.~\ref{fig:stellarmass}). Unfortunately, since the relation between
stellar mass and halo mass for central  galaxies is relatively narrow,
we  cannot discriminate   whether  this trend is   causally  linked to
stellar mass or to halo  mass. 

In  addition  to  the  {\it  occurrence} of   activity,  we have  also
investigated the halo  and stellar mass  dependencies  of the activity
{\it strengths}. The activity strength of star forming and optical AGN
centrals is seen to systematically decrease with both stellar and halo
mass by  a factor of 4  - 2  across  the range $10^{9.8} <  M_{\ast} <
10^{11.2} h^{-2} \Msun$  and  $10^{12} <  M_{\rm h} <   10^{14} h^{-1}
\Msun$, while the specific radio power  of HRA centrals increases by a
factor  of 1.5. Again,  the    tight relation between  $M_{\ast}$  and
$M_{\rm h}$ prevents   us from investigating  which is   more causally
connected to the activity strength.

In hierarchical models of  structure formation, dark matter haloes and
galaxies continuously  grow in mass.   We may therefore  interpret the
mass  dependence of  central galaxies  as a  trend with  time: central
galaxies start out forming stars,  but once their stellar mass exceeds
a few times  $10^{10} h^{-2} \Msun$ (or their  halo mass exceeds $\sim
10^{12} h^{-1}  \Msun$), star formation  is quenched, and  optical AGN
activity  becomes important.   By the  time  the halo  mass starts  to
exceed a few time $10^{13}  h^{-1} \Msun$ (or the stellar mass exceeds
$\sim 10^{11} h^{-2} \Msun$), the radio-mode of AGN activity starts to
become important,  while optical AGN activity is  somewhat suppressed. 
Interestingly, the mass scale  of this first transition coincides with
the  mass scale  where (i)  the  mass-to-light ratios  of dark  matter
haloes are  minimal (Yang,  Mo \&  van den Bosch  2003; van  den Bosch
\etal 2007),  (ii) dark  matter haloes start  to develop  an accretion
shock at their virial radius, giving rise to a hot corona (Birnboim \&
Dekel 2003; Keres  et al 2005; Birnboim, Dekel  \& Neistein 2007), and
(iii) galaxies undergo a transition from being disk-dominated to being
bulge-dominated  (Kauffmann  \etal   2003a).   

The following picture emerges: central galaxies in haloes with $M_{\rm
  h} \lta  10^{12} h^{-1}\Msun$ accrete cold gas,  which they actively
turn into stars.  During this period supernova feedback is believed to
play an important role (Dekel \&  Silk 1986) and to be responsible for
an increase of the baryonic mass fraction that is converted into stars
with increasing halo mass. If a galaxy undergoes a major merger during
this epoch,  it regrows  a disk around  its newly formed  bulge (e.g.,
Kauffmann,  White \&  Guiderdoni  1993).  The  tight relation  between
bulge mass and  black hole mass suggests that  this bulge formation is
somehow coupled to  the formation of a supermassive  black hole at its
center (Gebhardt et  al.  2000; Ferrarese \& Merritt  2000).  Once the
halo mass exceeds  $\sim 10^{12} h^{-1}\Msun$, it starts  to develop a
rapidly expanding  accretion shock, which heats up  any newly accreted
gas to  the halo's  virial temperature.  Somehow  most of this  gas is
prevented from cooling, resulting in a quenching of the star formation
of the  central galaxy (e.g.,  Dekel \& Birnboim 2006;  Cattaneo \etal
2006; Birnboim \etal 2007). From this point onwards, the galaxy cannot
regrow a  new disk after it  undergoes a major  merger, which explains
the transition towards more bulge dominated systems. These mergers are
also likely  to drive any  remaining gas to  the center of  the merger
remnant,  where it gives  rise to  a starburst  (exhaust the  cold gas
supply) and to AGN activity  (e.g. Mihos \& Hernquist 1996; Sanders \&
Mirabel 1996).  The latter may  play an important role in reheating or
expelling gas, and  thus in quenching the star  formation (e.g., Menci
\etal  2005; Springel,  Di  Matteo \&  Hernquist  2005; Hopkins  \etal
2006).  Once the halo has grown a hot corona that is massive and dense
enough (i.e., when  $M_{\rm h}$ starts to exceed  a few times $10^{13}
h^{-1} \Msun$), it starts to dissipate the energy output from the AGN,
giving rise  to radio  lobes, i.e.  the hot gas  serves as  a `working
surface' for the  jet (e.g. Kauffmann, Heckman \&  Best 2008). This in
turn gives rise to radio-mode  AGN feedback, which is believed to play
an important role  in preventing the hot gas  from cooling and forming
stars (e.g., Bower  \etal 2006; Croton \etal 2006;  Kang, Jing \& Silk
2006; Monaco, Fontanot \& Taffoni 2007).

As to satellite  galaxies,  our results support   a  picture in  which
satellite galaxies, in general, have their activity quenched.  This is
supported by the fact that the fraction of active satellites is always
smaller than the fraction of active centrals of  the same stellar mass
(see    Figs.~\ref{fig:stellarmass}),  and  by     the  fact  that the
probability for an active galaxy to be a central galaxy is higher than
that  for    random   galaxy   of     the  same  stellar    mass  (see
Fig.~\ref{fig:censat}).  The  fact that satellite galaxies only reveal
a weak  dependence of activity on halo  mass (with more massive haloes
showing somewhat  smaller  fractions of active  satellites), indicates
that the efficiency of  this activity-shutdown mechanism must  operate
with roughly equal  efficiency in haloes of all   masses.  This is  in
good agreement with van den Bosch \etal (2008a,b)  who, using the same
galaxy   group catalogue as that    used  here, found that   satellite
galaxies are redder and more concentrated than central galaxies of the
same stellar mass, but that the magnitude of  this effect is virtually
independent of halo mass. As discussed in van den Bosch \etal (2008a),
this seems   to  favor  strangulation  (i.e.,   the  removal  of   the
satellite's hot gas reservoir) as the main quenching mechanism, rather
than processes such as  ram-pressure stripping which mainly operate in
the dense ICM of massive clusters.

The  activity  of satellite  galaxies  shows  a  strong dependence  on
stellar  mass, very  similar to  that  of central  galaxies.  This  is
consistent  with a picture  in which  satellite galaxies  were central
galaxies prior to being accreted, thus `inheriting' their stellar mass
dependence. We  also find that  the activity of satellite  galaxies is
correlated  with their  projected group-centric  distance:  while star
forming satellites  preferentially reside in  the outskirts, satellite
galaxies with optical AGN activity have smaller projected distances to
their group centers than  the average satellite. However, these trends
do not  reflect a causal connection between  activity and halo-centric
distance.  Rather,  they are simply  a reflection of  mass segregation
(more massive satellites have smaller halo-centric distances) combined
with the relation between activity  and stellar mass. The absence of a
direct link between activity  and halo-centric distance indicates that
the activity-quenching mechanism(s) must operate on time scales longer
than,  or comparable to  the dynamical  time in  the haloes.   This is
again consistent  with the idea  that the main quenching  mechanism is
strangulation, which seems  to operate on a timescale  of several Gyrs
(e.g. McCarthy  \etal 2008; Kang \&  van den Bosch 2008),  and is also
supported by the lack of  correlation between the activity strength of
satellites and  either stellar or halo  mass as shown in  Figs. 11, 12
and 13.

%


\end{document}